\documentclass[aps,prd,twocolumn,amsmath,superscriptaddress,amssymb,showpacs,floatfix,nofootinbib]{revtex4-1}

\usepackage{hyperref}
\usepackage{graphicx}
\usepackage{dcolumn}
\usepackage{xcolor}
\usepackage{bm}
\usepackage{natbib}
\usepackage{calligra}
\usepackage[T1]{fontenc}
\usepackage{egothic}
\usepackage[T1]{fontenc}
\newfont{\rsfsten}{rsfs10 scaled 1200}
\newfont{\rsfsseven}{rsfs10 scaled 1200}
\newfont{\rsfsfive}{rsfs10 scaled 1200}
\usepackage{epsfig}
\usepackage{units}
\usepackage[utf8]{inputenc}

\newcommand{\be}{\begin{equation}}
\newcommand{\ee}{\end{equation}}
\newcommand{\bea}{\begin{eqnarray}}
\newcommand{\eea}{\end{eqnarray}}

\def\lsim{\mathrel{\raise.3ex\hbox{$<$\kern-.75em\lower1ex\hbox{$\sim$}}}}
\def\gsim{\mathrel{\raise.3ex\hbox{$>$\kern-.75em\lower1ex\hbox{$\sim$}}}}

\begin{document}
\hspace{13cm} \parbox{5cm}{FERMILAB-PUB-20-271-AE-T}

\hspace{13cm}
\vspace{0.6cm}

\title{Constraining the Charge-Sign and Rigidity-Dependence of Solar Modulation}

\author{Ilias Cholis}
\email{cholis@oakland.edu, ORCID: orcid.org/0000-0002-3805-6478}
\affiliation{Department of Physics, Oakland University, Rochester, Michigan, 48309, USA}
\author{Dan Hooper}
\email{dhooper@fnal.gov, ORCID: orcid.org/0000-0001-8837-4127}
\affiliation{Fermi National Accelerator Laboratory, Center for Particle Astrophysics, Batavia, Illinois, 60510, USA}
\affiliation{University of Chicago, Department of Astronomy and Astrophysics, Chicago, Illinois, 60637, USA}
\affiliation{University of Chicago, Kavli Institute for Cosmological Physics, Chicago, IL 60637, USA}
\author{Tim Linden}
\email{linden.70@osu.edu, ORCID: orcid.org/0000-0001-9888-0971}
\affiliation{Stockholm University and the Oskar Klein Centre, Stockholm, Sweden}

\date{\today}

\begin{abstract}

Our ability to identify the sources of cosmic rays and understand how these particles propagate through the interstellar medium is hindered by the combined effects of the solar wind and its embedded magnetic field, collectively known as solar modulation. In this paper, we build upon our previous work to model and constrain the effects of solar modulation on the cosmic-ray spectrum, using data from {\textit AMS-02} and BESS Polar II collected between 2007 and 2012, during which the heliospheric magnetic field was in a state of negative polarity. Our model uses measurements of the heliospheric magnetic field and the tilt angle of the heliospheric current sheet to accurately predict the effects of solar modulation as a function of time, charge, and rigidity. By incorporating data from a period of negative polarity, we have been able to robustly observe and constrain the charge-dependent effects of solar modulation.

\end{abstract}

\maketitle

\section{Introduction}
\label{sec:introduction}

At energies up to the PeV-scale, the cosmic-ray spectrum is dominated by particles that are produced by galactic sources, such as supernova remnants and pulsars, as well as through the interactions of primary cosmic rays with the interstellar medium (ISM). The spectrum of these particles in the local ISM varies little with time. This is because they originate from a large number of sources, and most of the volume of the ISM maintains relatively stable properties (as far as cosmic-ray propagation is concerned) on the timescales of Galactic cosmic-ray propagation, typically on the order of $\sim$10 Myr or less. As they enter the heliosphere that surrounds the Solar System, however, cosmic rays are subject to the effects of the solar wind and its embedded magnetic field, which are observed to vary significantly over timescales of months and years. The amplitude of the local heliospheric magnetic field has even been observed to change by as much as a factor of two over the course of a single day. 

At GeV energies, it takes cosmic rays anywhere between a few months to about a year to travel from the edge of the heliosphere to the vicinity of Earth. The effects of the solar wind and the heliospheric magnetic field, collectively known as solar modulation, are highly time-dependent and can be clearly observed in the measured fluxes of cosmic rays with rigidities \mbox{$R \equiv |p|/q \lsim 10\,{\rm GV}$} (where $p$ and $q$ are the momentum and charge of the cosmic rays). In this same energy range, the situation is significantly complicated by the effects of diffusion, diffusive reacceleration, convective winds, and electron/positron energy loss processes in the ISM. In light of these factors, it has been challenging to reliably study the detailed effects of solar modulation on the cosmic-ray spectrum. In this paper, we build upon our previous work~\cite{Cholis:2015gna} (see also Refs.~\cite{Corti:2019jlt,Boschini:2019ubh,Wang:2019xtu,Gieseler:2017xry,Corti:2018ycg,Aslam:2018kpi,Bertucci:2019ypo}) in an effort to isolate and predictively model the effects of solar modulation on the cosmic-ray spectrum.  

The impact of solar modulation on the cosmic-ray spectrum is given in terms of the potential, $\Phi$, which allows us to relate the spectrum at Earth to that present in the local ISM~\cite{1968ApJ...154.1011G}:
\begin{eqnarray}
\frac{dN^{\oplus}}{dE_{\rm kin}} (E_{\rm kin}) &=& \frac{(E_{\rm kin}+m)^{2} -m^{2}}{(E_{\rm kin}+m+ |Z| e \Phi)^{2} -m^2} \nonumber \\ 
&\times&\; \frac{dN^{\rm ISM}}{dE_{\rm kin}} (E_{\rm kin}+ | Z | e \Phi),
\end{eqnarray}
where $E_{\rm kin}$ is the kinetic energy of the particle at Earth, $E_{\rm kin}+ | Z | e \Phi$ is its kinetic energy in the ISM, and the labels ``ISM'' and ``$\oplus$'' denote values in the local ISM and at the location of Earth, respectively. The quantities $m$ and $|Z|e$ represent the mass and absolute charge of the cosmic rays.

In a previous study, we used measurements of the cosmic-ray proton spectrum from \textit{PAMELA}, \textit{AMS-02}, \textit{Voyager 1} and a number of balloon-based experiments from the 1990s and the 2000s, to produce a predictive, time-, charge- and rigidity-dependent formula for the solar modulation potential~\cite{Cholis:2015gna}:
 \begin{eqnarray}
\Phi(R,q,t) = &\phi_{0}& \, \bigg( \frac{|B_{\rm tot}(t)|}{4\, {\rm nT}}\bigg) + \phi_{1} \, H(-qA(t))\, \bigg( \frac{|B_{\rm tot}(t)|}{4\,  {\rm nT}}\bigg) \nonumber \\
&\times& \,\bigg(\frac{1+(R/R_0)^2}{\beta (R/R_{0})^3}\bigg) \, \bigg( \frac{\alpha(t)}{\pi/2} \bigg)^{4},
\label{eq:ModPot}
\end{eqnarray}
where $|B_{\rm tot}|$ and $A$ are the strength and polarity of the heliospheric magnetic field (as measured at Earth), and $\alpha$ is the tilt angle of the heliospheric current sheet, which separates the north and south magnetic hemispheres. These quantities are each time-dependent, and can be measured independently of the cosmic-ray spectrum. $R$, $q$ and $\beta$ are the rigidity, charge and velocity (as evaluated in the local ISM) of the cosmic ray, respectively, and $R_0$ is a reference rigidity which we take to be $0.5 \, {\rm GV}$. $H$ is the Heaviside step function, which is equal to zero or unity depending on the sign of $-qA$. 

During periods of positive polarity, positively charged cosmic rays are able to travel relatively efficiently to Earth, experiencing only modest energy losses (the same is true for negatively charged particles during periods of negative polarity). In contrast, in situations where $qA < 0$, cosmic rays lose significantly more energy as they propagate from the heliopause into the Inner Solar System along the heliospheric current sheet. Furthermore, while low-rigidity cosmic rays are confined to travel along the twisting heliospheric current sheet, higher-rigidity particles are capable of drifting across the sheet (see Refs.~\cite{2012Ap&SS.339..223S, Potgieter:2013pdj, Cholis:2015gna} for further discussion), increasing the efficiency of transport into the inner heliosphere and leading to the rigidity-dependent effect that is represented by the second term in Eq.~\ref{eq:ModPot}.

\begin{figure}
\hspace{-0.15in}
\vspace{-0.3in}
\includegraphics[width=3.55in,angle=0]{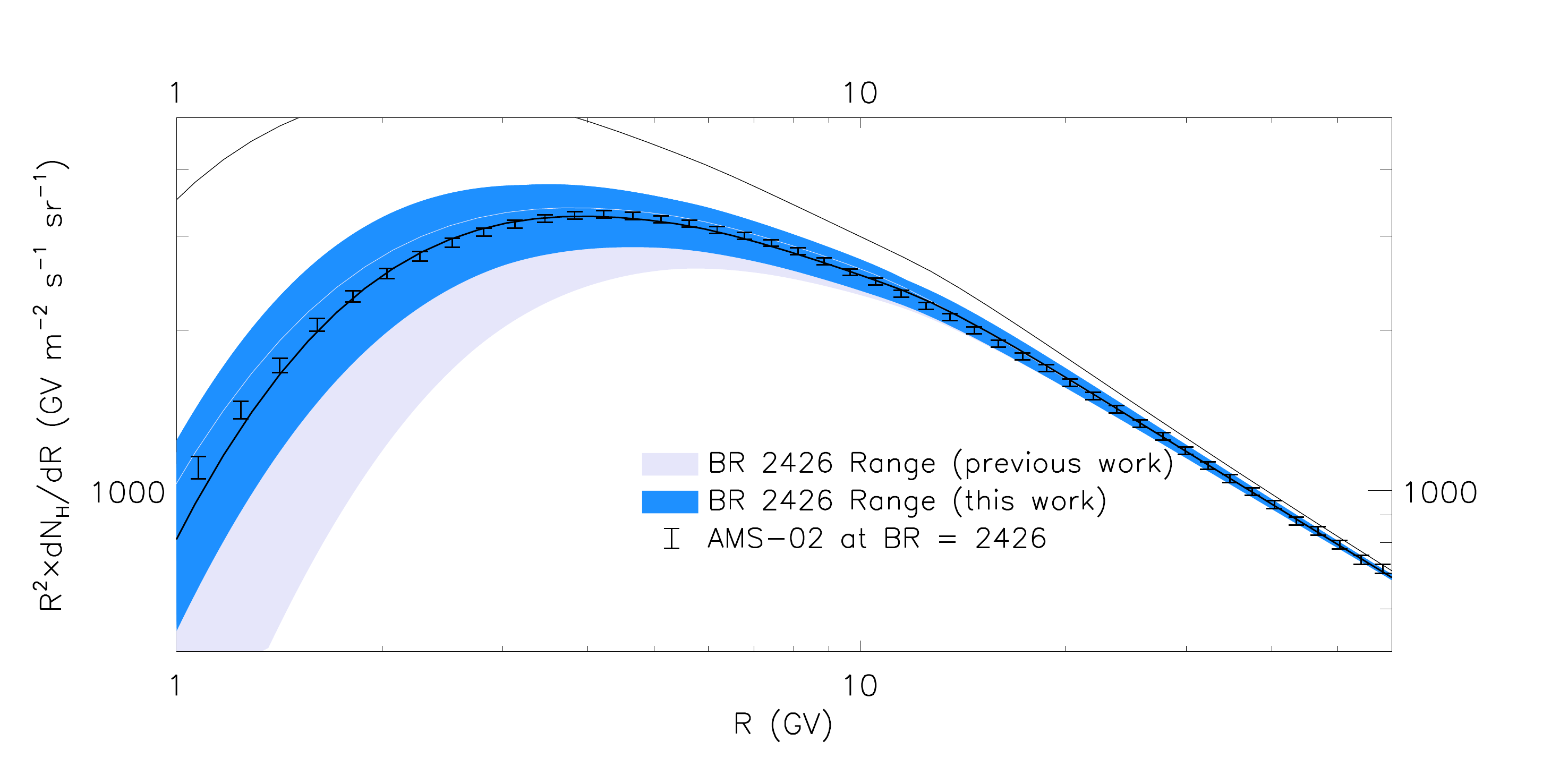} \\
\hspace{-0.15in}
\vspace{-0.3in}
\includegraphics[width=3.55in,angle=0]{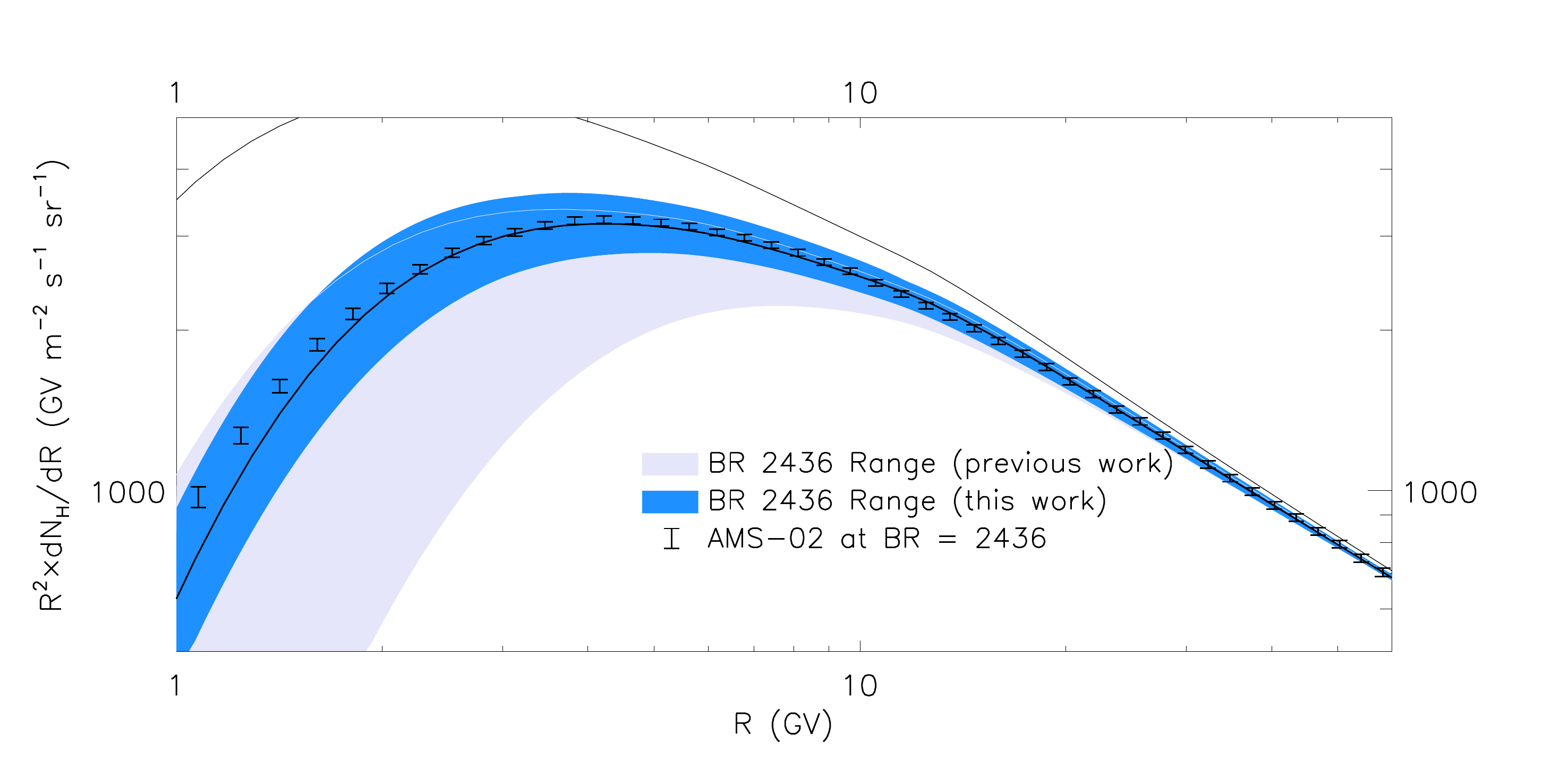} \\
\hspace{-0.15in}
\vspace{-0.3in}
\includegraphics[width=3.55in,angle=0]{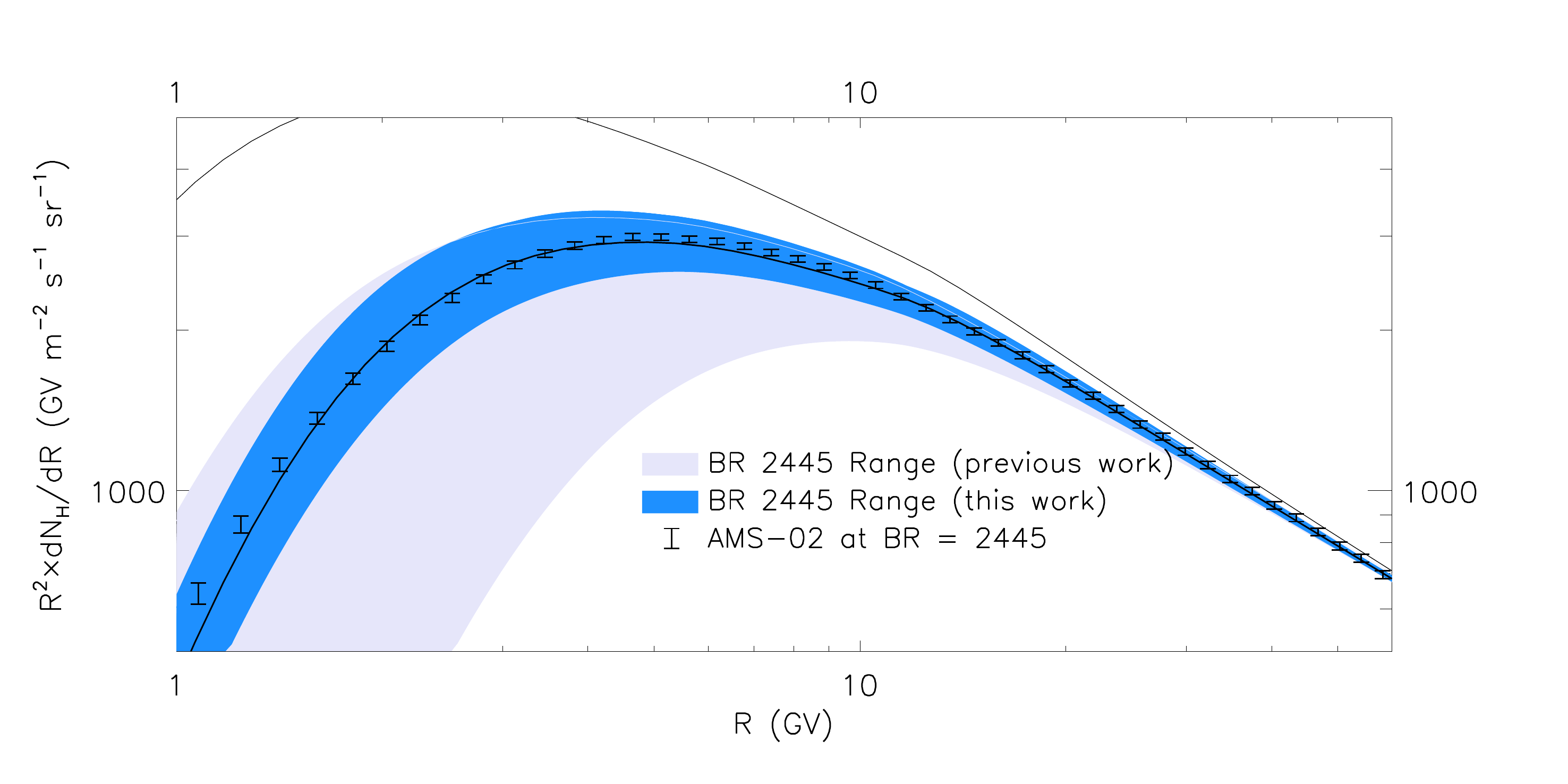}
\vspace{0.05in}
\caption{The cosmic-ray hydrogen spectrum (including both protons and deuterons) as measured by \textit{AMS-02} during three representative Bartels' Rotations (approximately 27-day periods, associated with the apparent rotation of the Sun about its axis). These Bartels' Rotations (2426, 2436 and 2445) are associated with periods in May-June 2011, February-March 2012, and October 2012.
In each frame, the wider bands  (light blue) represent the predictions of Eq.~\ref{eq:ModPot}, including the uncertainties associated with the impact of solar modulation, using the determination of $\phi_0$ and $\phi_1$ from our previous work~\cite{Cholis:2015gna}. The narrower bands (dark blue) represent the significantly smaller uncertainty ranges based on the determinations of $\phi_0$ and $\phi_1$ obtained in this study. Also shown are the central predictions of the spectrum, both before (above the bands) and after (near the center of each band) accounting for the effects of solar modulation. For the central values of the modulated spectra, we have adopted $\phi_{0} = 0.315$ GV and $\phi_{1} = 1.5$ GV.}
\label{fig:HydrogenSpect}
\end{figure}

In our previous study, we found that the combined data supported a value of $ \phi_{0} = 0.32-0.38 \, {\rm GV}$ (at $2\sigma$ confidence), while $\phi_{1}$ was only very weakly constrained to lie between zero and 16 GV~\cite{Cholis:2015gna}. This model (with the above mentioned values of $\phi_{0}$ and $\phi_{1}$), allows one to use the measured quantities of $|B_{\rm tot}|$, $A$, and $\alpha$ to predict the value of the solar modulation potential as a function of time. As can be seen in Fig.~\ref{fig:HydrogenSpect}, however, there are large uncertainties associated with the predictions of this model. In this figure, we show the cosmic-ray hydrogen spectrum (which includes both protons and deuterons) as measured by \textit{AMS-02} during three representative Bartels' Rotations (approximately 27-day periods, associated with the apparent rotation of the Sun about its axis). These measurements are compared to the spectrum predicted by our model during each Bartels' Rotation, including light blue bands which represent the uncertainties associated with the impact of solar modulation, as presented in Ref.~\cite{Cholis:2015gna}. Although the data is consistently in good agreement with the model, the uncertainties are often quite large. It is the primary goal of this work to reduce these uncertainties and, more specifically, to reduce the uncertainties on the parameters $ \phi_{0}$ and $ \phi_{1}$. The smaller dark blue bands shown in Fig.~\ref{fig:HydrogenSpect} represent the significantly smaller uncertainties on the impact of solar modulation, based on the determinations of $\phi_0$ and $\phi_1$ obtained in this study.

\begin{figure*}
\begin{centering}
\hspace{-0.15in}
\includegraphics[width=5.55in,angle=0]{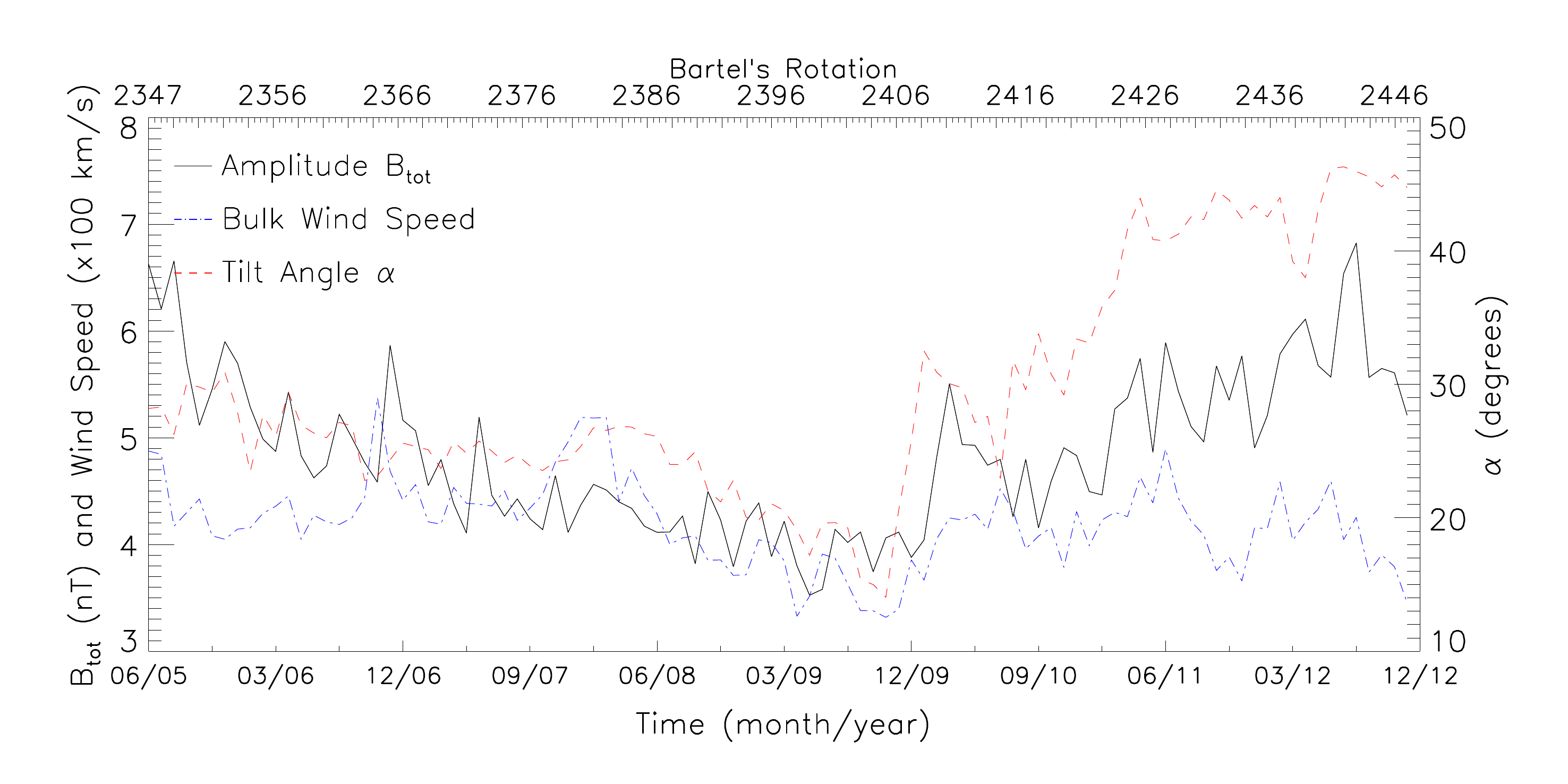}
\end{centering}
\caption{The amplitude of the heliospheric magnetic field, $B_{\rm tot}$, the tilt angle of the heliospheric current sheet, $\alpha$, and the bulk speed of the Solar Wind, in the period between 2006 and 2012. The measurements of $B_{\rm tot}$ and the Solar Wind speed are from the Advanced Composition Explorer magnetometer~\cite{ACESite}, while the determinations of the tilt angle are based on the public model of the Wilcox Solar Observatory~\cite{WSOSite}.}
\label{fig:HMF_vs_time}
\end{figure*}

\section{Methodology}
\label{sec:method}

In order to test the solar modulation model described by Eq.~\ref{eq:ModPot}, and to further constrain the values of the parameters $\phi_{0}$ and $\phi_{1}$, we focus on the period of time preceding the last polarity flip of the heliospheric magnetic field, which started at the end of 2012. To this end, we make use of two data sets. First, we utilize data from the \textit{AMS-02} experiment onboard the International Space Station, which includes time-dependent measurements of the cosmic-ray hydrogen flux at rigidities between 1.16 and 10.1 GV~\cite{Consolandi:2015sop}. Second, we make use of BESS Polar II measurements of the antiproton-to-proton ratio, $\bar{p}/p$. For BESS Polar II, these measurements are of the cosmic-ray spectrum between 0.64 and 4.1 GV, and were taken in December 2007~\cite{Abe:2017yrg}. 

The predictive power of Eq.~\ref{eq:ModPot} comes from our ability to independently measure the values of 
$B_{\rm tot}$, $A$ and $\alpha$. From the values of these quantities, we can directly calculate 
$\Phi$ for any cosmic-ray species, as a function of rigidity and time. 
In Fig.~\ref{fig:HMF_vs_time}, we show the time evolution of the heliospheric magnetic field amplitude, $B_{\rm tot}$, as measured by the Advanced Composition Explorer (\textit{ACE}) magnetometer~\cite{ACESite}, as well as the tilt angle of the heliospheric current sheet, $\alpha$, as given by the public model of the 
Wilcox Solar Observatory~\cite{WSOSite}.  For comparison, we also plot the bulk wind speed, although this quantity is not directly used in our solar modulation model.

During periods of negative polarity (including the era considered in this study), simulations have shown that negatively charged particles travel from the outer parts of the heliosphere to the vicinity of Earth through the poles of the heliospheric magnetic field, experiencing a typical travel time of a few months. In contrast, positively charged particles travel predominantly along the heliospheric current sheet, which simulations show requires about a year for a $\sim$\,$100 \, {\rm MeV}$ proton (see, for example, Ref.~\cite{2012Ap&SS.339..223S}). This suggests that the values of $B_{\rm tot}$ and $\alpha$ that are used in Eq.~\ref{eq:ModPot} should reflect some sort of average over a period of months or years. With this issue in mind, we have tested a variety of time-averaging schemes for $B_{\rm tot}$ and $\alpha$, including averaging over different time windows and with different weighting procedures.

\begin{table*}[t]
    \begin{tabular}{ccccccccccc}
         \hline
           Model & $\delta$ & $z_{L}$ (kpc) & $D_{0} \times 10^{28}$ (cm$^2$/s) & $v_{A}$ (km/s) & $dv_{c}/dz$ (km/s/kpc) & $\alpha_{1}$ & $R_{br_{1}}$ (GV) & $\alpha_{2}$ & $R_{br_{2}}$ (GV) & $\alpha_{3}$\\
            \hline \hline
            A  & 0.33 & 6.0 & 6.50 & 30.0 & 0 & 1.75 & 6.0 & 2.04 & 14.0 & 2.41 \\
            B &  0.37 & 5.5 & 5.50 & 30.0 & 2.5 & 1.72 & 6.0 & 2.00 & 12.4 & 2.38 \\
            C &  0.40 & 5.6 & 4.85 & 24.0 & 1 & 1.69 & 6.0 & 2.00 & 12.4 & 2.38 \\
            D &  0.45 & 5.7 & 3.90 & 25.7 & 6 & 1.69 & 6.0 & 1.99 & 12.4 & 2.355 \\ 
            E &  0.50 & 6.0 & 3.10 & 23.0 & 9 & 1.71 & 6.0 & 2.02 & 11.2 & 2.38 \\
            F &  0.40 & 3.0 & 2.67 & 22.0 & 3 & 1.68 & 6.0 & 2.08 & 13.0 & 2.41 \\
            \hline \hline 
        \end{tabular}
    \caption{The six models used in this study to describe the injection and propagation of cosmic rays in the interstellar medium (ISM). These models are each in good agreement with the measured boron-to-carbon ratio, proton spectrum, and a variety of other cosmic-ray data.}
    \label{tab:ISM_models}
\end{table*}

\begin{figure*}
\vskip -0.25in
\includegraphics[width=3.5in,angle=0]{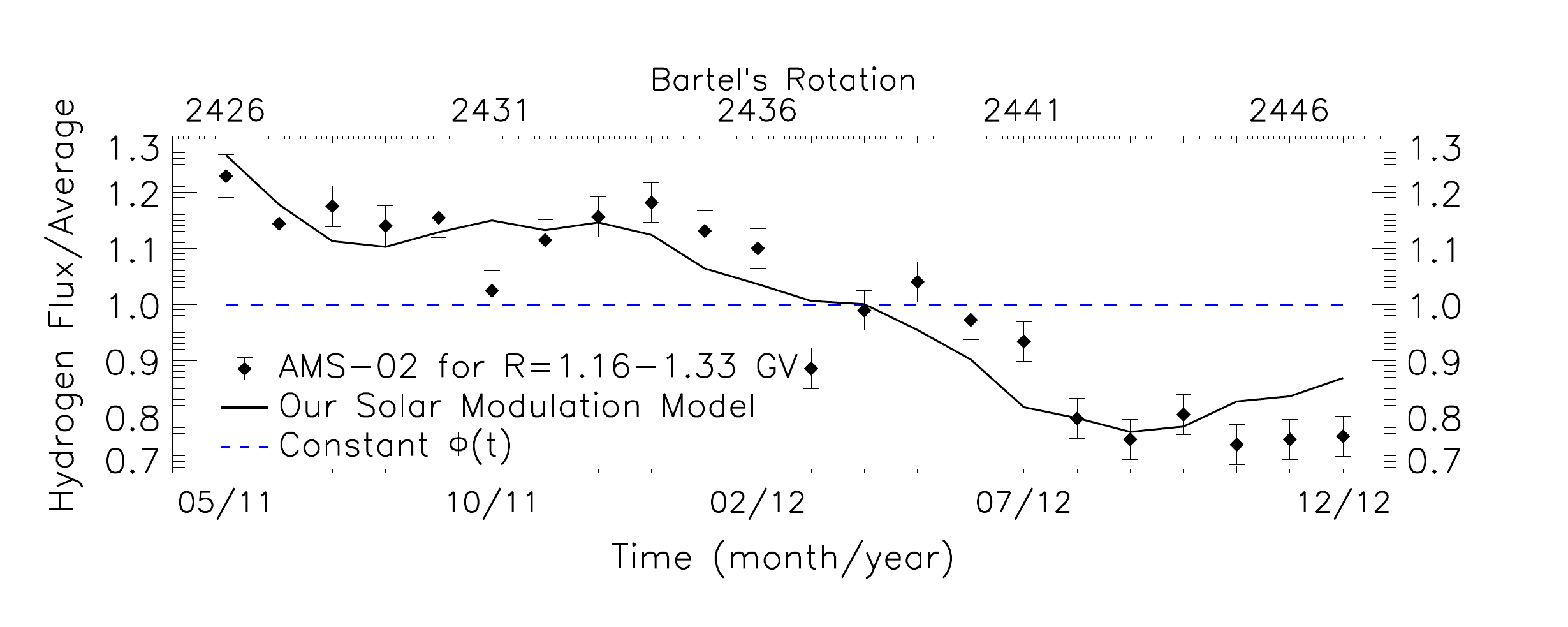}
\includegraphics[width=3.5in,angle=0]{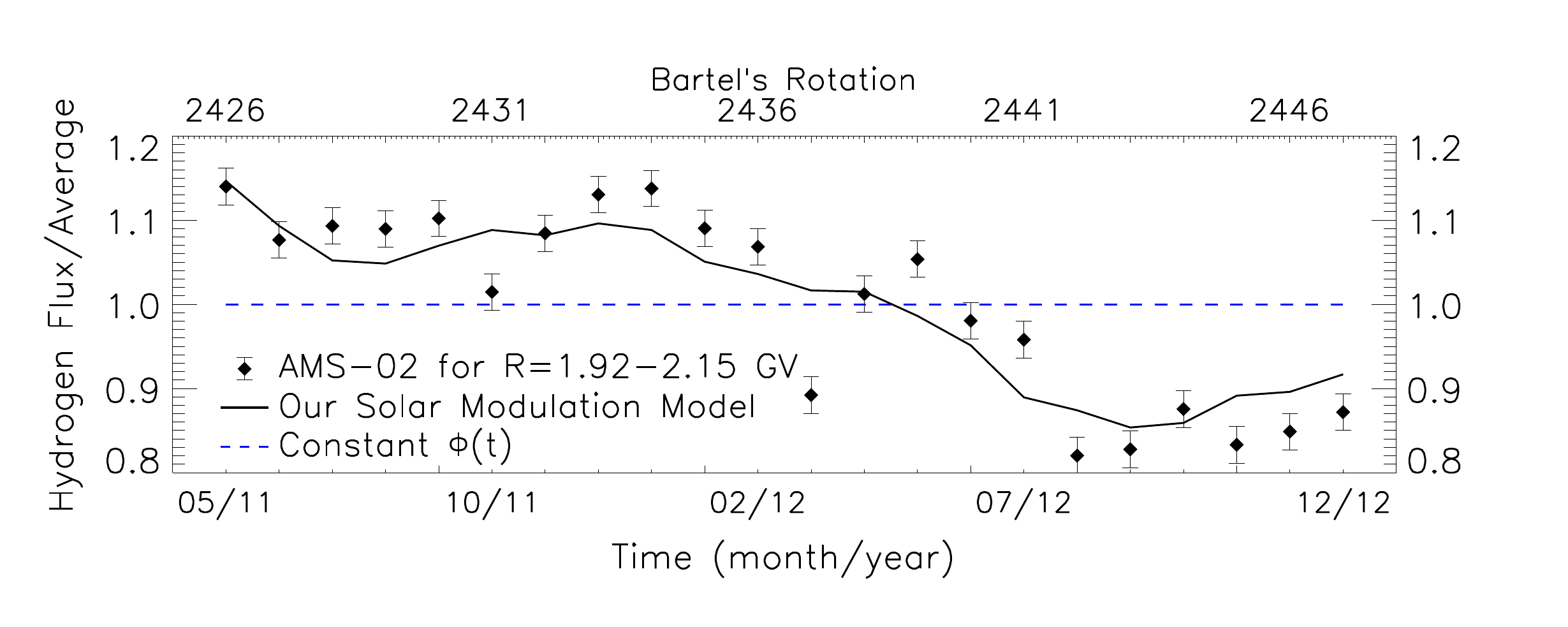}\\
\vskip -0.15in
\includegraphics[width=3.5in,angle=0]{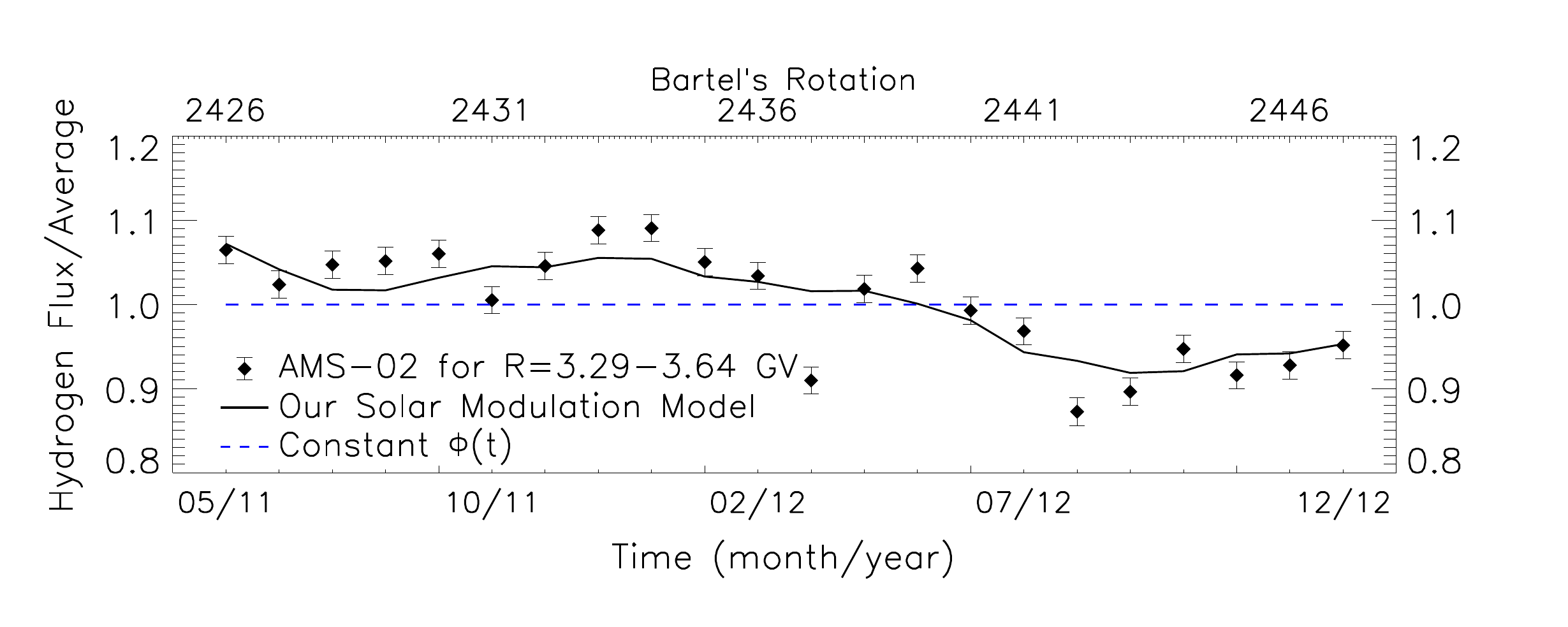}
\includegraphics[width=3.5in,angle=0]{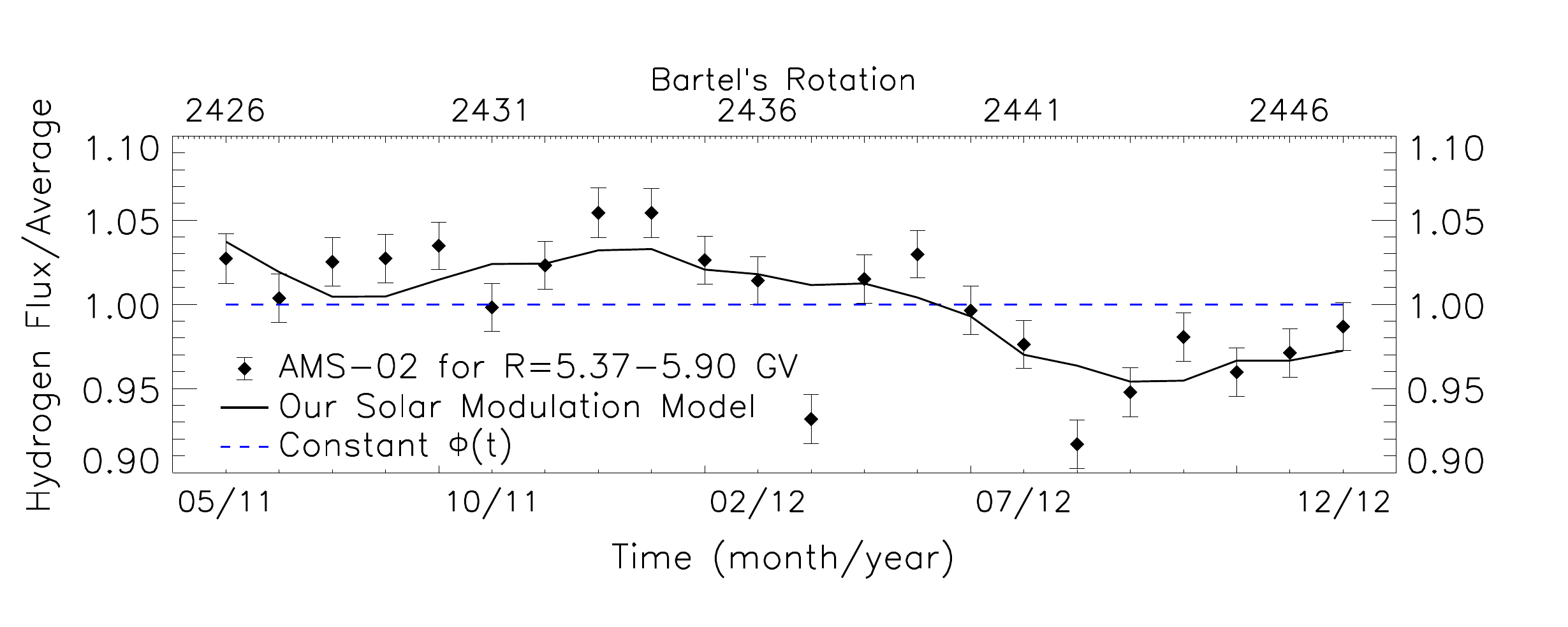}\\
\vskip -0.15in
\includegraphics[width=3.5in,angle=0]{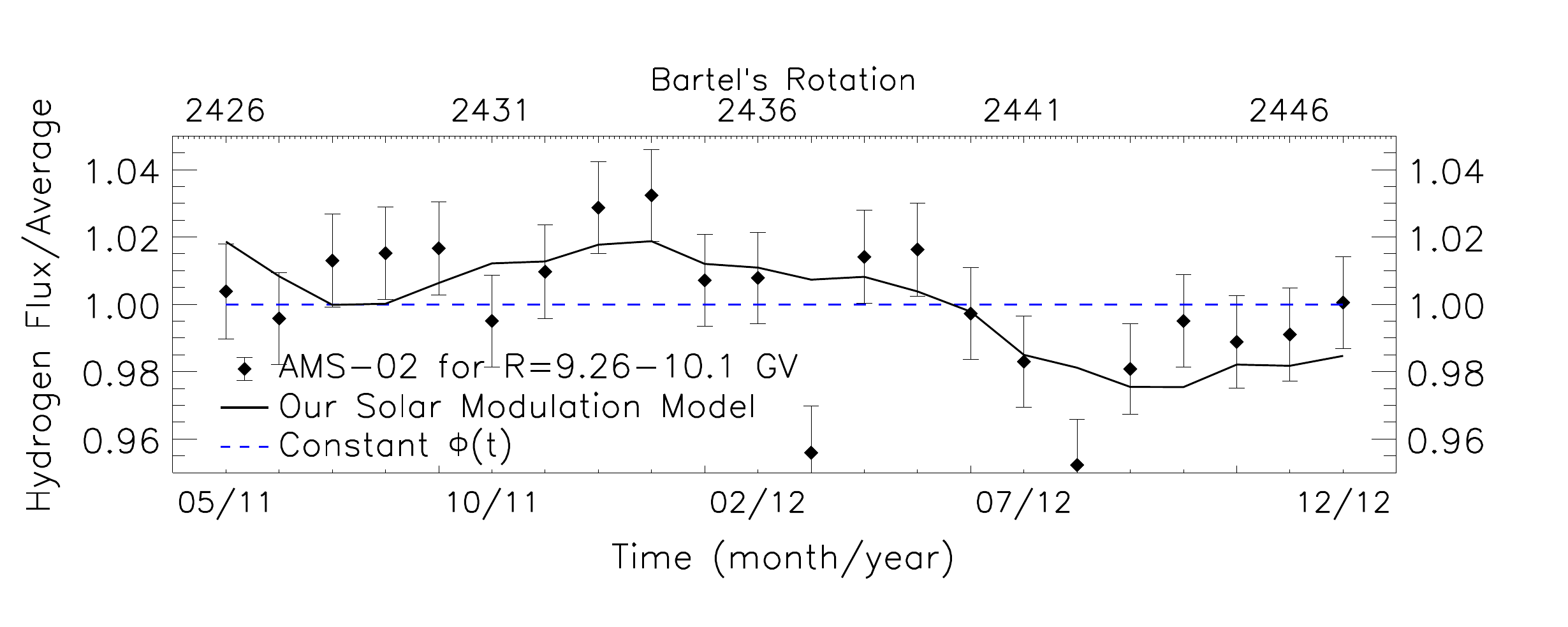}
\vskip -0.2in
\caption{The time variation of the cosmic-ray hydrogen flux (including both protons and deuterons) in five ranges of rigidity, as measured by~\textit{AMS-02} between May 2011 and December 2012 (Bartels' Rotations 2426 to 2447). We compare these measurements to the predictions of our solar modulation model (see Eq.~\ref{eq:ModPot}), for the best-fit values of $\phi_0=0.315 \, {\rm GV}$ and $\phi_1=1.5 \, {\rm GV}$. Our model provides a good overall description of the time-evolution and rigidity-dependence of the effects of solar modulation.}
\label{fig:HRatio}
\end{figure*}  

To account for the systematic uncertainties associated with the cosmic-ray spectrum in the local ISM, we have employed the six models described in Table~\ref{tab:ISM_models}. These models each make different assumptions regarding the injected spectrum of cosmic rays, as well as their diffusion, convection, and diffusive reacceleration in the galactic medium. We model the injection and propagation of cosmic rays through the Galaxy using \texttt{Galprop v56}, which
 numerically solves the transport equation to determine the flux of primary and secondary cosmic-ray 
 species in the local ISM~\cite{GALPROPSite, Strong:2015zva, NEWGALPROP, galprop}.  
These six models each provide a good fit to the boron-to-carbon ratio as measured by \textit{AMS-02} and \textit{PAMELA}, the cosmic-ray proton spectrum as measured by \textit{Voyager 1} and~\textit{AMS-02}, and measurements from \textit{AMS-02} of the carbon spectrum, helium spectrum, and the antiproton-to-proton ratio. To minimize the impact of the ISM model on our results, we have also made use of cosmic-ray ratios which cancel most of the relevant uncertainties. As we will show in the Sec.~\ref{sec:results}, our conclusions regarding solar modulation are not significantly impacted by our choice of ISM model. Unless stated otherwise, the results shown in this paper are calculated using model C for the injection and transport of cosmic rays in the ISM. For the cosmic-ray antiproton-to-proton ratio, $\bar{p}/p$, there are also significant uncertainties associated with the antiproton production cross section in inelastic proton-proton collisions. We treat these uncertainties following the procedures described in our previous work~\cite{Cholis:2017qlb, Cholis:2019ejx}.

\section{Results}
\label{sec:results}

In Fig.~\ref{fig:HRatio}, we plot the time variation of the cosmic-ray hydrogen (including both protons and deuterons) flux in five ranges of rigidity, as measured by~\textit{AMS-02}. These results are shown for a period of time between May 2011 (when \textit{AMS-02} was installed on the International Space Station) and December 2012 (when the polarity flip of the heliospheric magnetic field began), consisting of 22 Bartels' Rotations (2426 to 2447). We compare these measurements to the predictions of our solar modulation model (see Eq.~\ref{eq:ModPot}), for the best-fit values of $\phi_0=0.315 \, {\rm GV}$ and $\phi_1=1.5 \, {\rm GV}$. For the values of $B_{\rm tot}$ and $\alpha$, we use the time-averaging scheme that was found to provide the best fit to the data. In particular, we take an average of $B_{\rm tot}$ over a period of 4 Bartels' Rotations, the most recent of which is 16 rotations prior to the time of the measurement. For $\alpha$, we take an average over the past 20 Bartels' Rotations. This is discussed further in Appendix~\ref{sec:AppendixB}, where we show results for a variety of other averaging schemes for $B_{\rm tot}$ and $\alpha$.

Time delays of this magnitude are not surprising given that the values of $B_{\rm tot}$ that we input into our model are those at the location of Earth, and it takes at least half a year for the magnetic field to propagate to the edge of the heliosphere (at $\sim\,100$ AU). These results suggest that cosmic-ray protons traveling through the heliospheric current sheet may take about a year to reach the inner Solar System from the heliosphere. Furthermore, the observed time delay suggests that the effects of solar modulation may be most significantly influenced by the amplitude of the magnetic field in the outer volume of the heliosphere, and thus most of the energy losses under consideration occur in the early stages of cosmic-ray propagation through the heliosphere.

The results shown in Fig.~\ref{fig:HRatio} demonstrate that our model predicts the time- and rigidity-dependence of the effects of solar modulation quite well. Compared to a constant flux of cosmic-ray hydrogen, our model (with $\phi_0=0.315 \, {\rm GV}$, $\phi_1=1.5 \, {\rm GV}$) is preferred at a level corresponding to $\Delta \chi^{2} =1160$. Furthermore, although we included a set of additional nuisance parameters in the form of a free normalization coefficient for each of the five rigidity ranges, the fit prefers values for these parameters that are each within $1\%$ of unity, demonstrating that our model (without these additional parameters) accurately describes the rigidity-dependence of solar modulation. 

Despite the successes of our model, there are times in which it significantly under-predicts the degree of rapid change in the cosmic-ray hydrogen flux. This is particularly true for the periods corresponding to Bartles' Rotations 2431, 2434, 2437, 2439 and 2442 (the 6th, 9th, 12th, 14th and 17th error bar from the left in Fig.~\ref{fig:HRatio}). These features are not easily explained within the context of our model, and may be indicative of secondary effects that cannot be directly inferred from the  measured values of $B_{\rm tot}$ and $\alpha$.

\begin{figure}
\begin{centering}
\includegraphics[width=3.25in,angle=0]{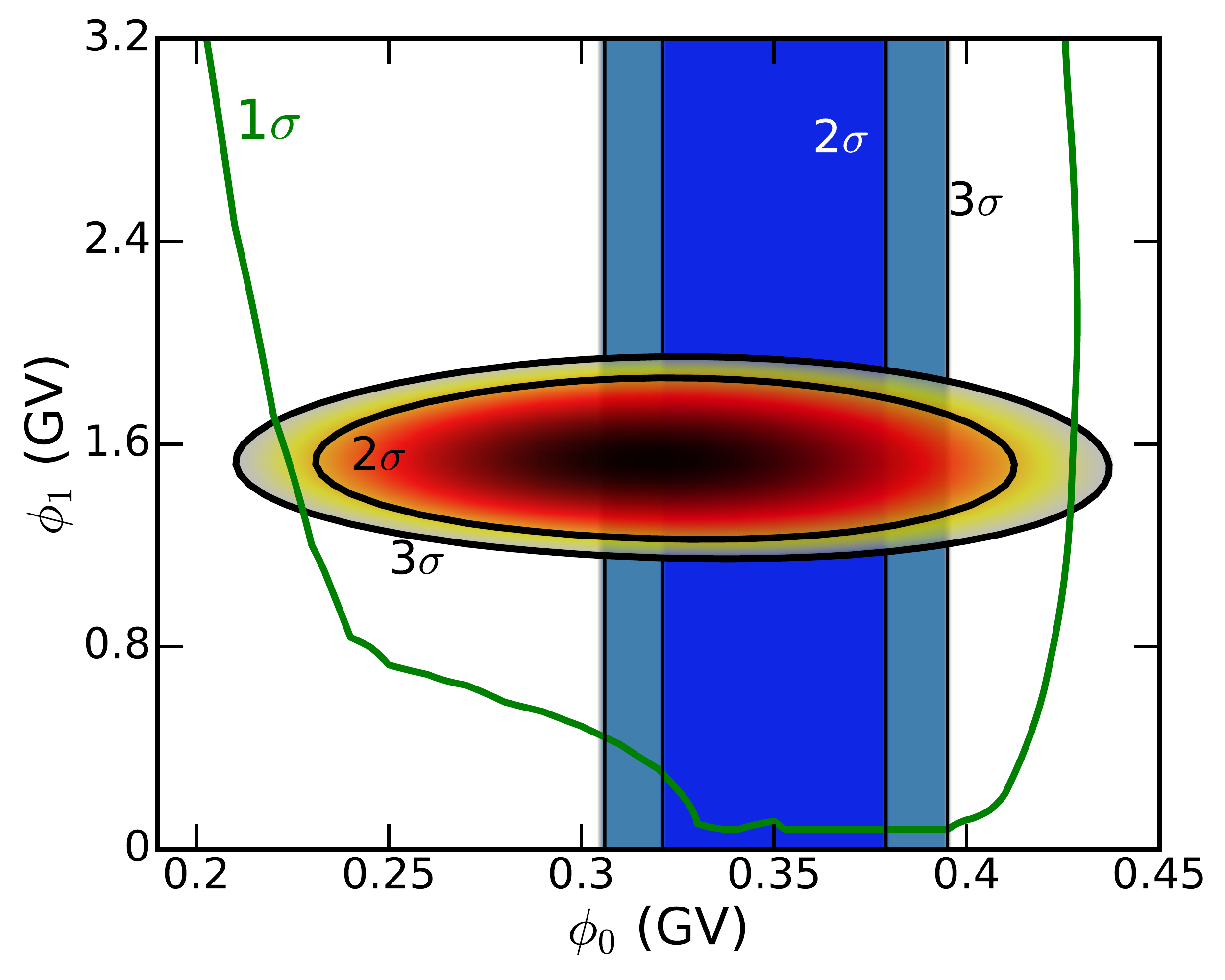}
\end{centering}
\caption{In the heat map, we show the constraints on $\phi_{0}$ and $\phi_{1}$, as derived in this study from the cosmic-ray hydrogen spectrum (including protons and deuterons) as measured by \textit{AMS-02}. At the $3\sigma$ level, these parameters are constrained to $\phi_0 = 0.210-0.435\,{\rm GV}$ and $\phi_1=1.15-1.95\,{\rm GV}$. The blue regions represent the constraints presented in our previous work~\cite{Cholis:2015gna}, which only weakly restricted the value of $\phi_1$. The green contour represents the region favored by the measurements of the antiproton-to-proton ratio, $\bar{p}/p$, by BESS Polar II and \textit{AMS-02}.}
\label{fig:PhiRanges}
\end{figure}

In Fig.~\ref{fig:PhiRanges}, we show the values of $\phi_{0}$ and $\phi_{1}$ that are preferred by the \textit{AMS-02} hydrogen data utilized in this study. At the $3\sigma$ level, this data allows us to constrain $\phi_0 = 0.210-0.435\,{\rm GV}$ and $\phi_1=1.15-1.95\,{\rm GV}$. In our previous work~\cite{Cholis:2015gna}, we produced a similar determination of $\phi_0 = 0.305-0.395\,{\rm GV}$, but were only able to place a relatively weak constraint on the charge-sign dependent effects of solar modulation, $\phi_1 \lsim 16\,{\rm GV}$ (at the 95\% confidence level). In contrast, we have now been able to strongly confirm the presence of these charge-sign dependent effects, excluding a value of $\phi_{1} = 0$ at a level of $\Delta \chi^{2}=275$ (corresponding to $\sim 16 \sigma$ significance) for the best-fit 16 Bartles' Rotation time delay, and $\Delta \chi^{2}=108$ ($\sim 11 \sigma$) when the time delay is treated as a free parameter (for additional details, see Appendix~\ref{sec:AppendixB}).

\begin{figure}
\begin{centering}
\includegraphics[width=3.25in,angle=0]{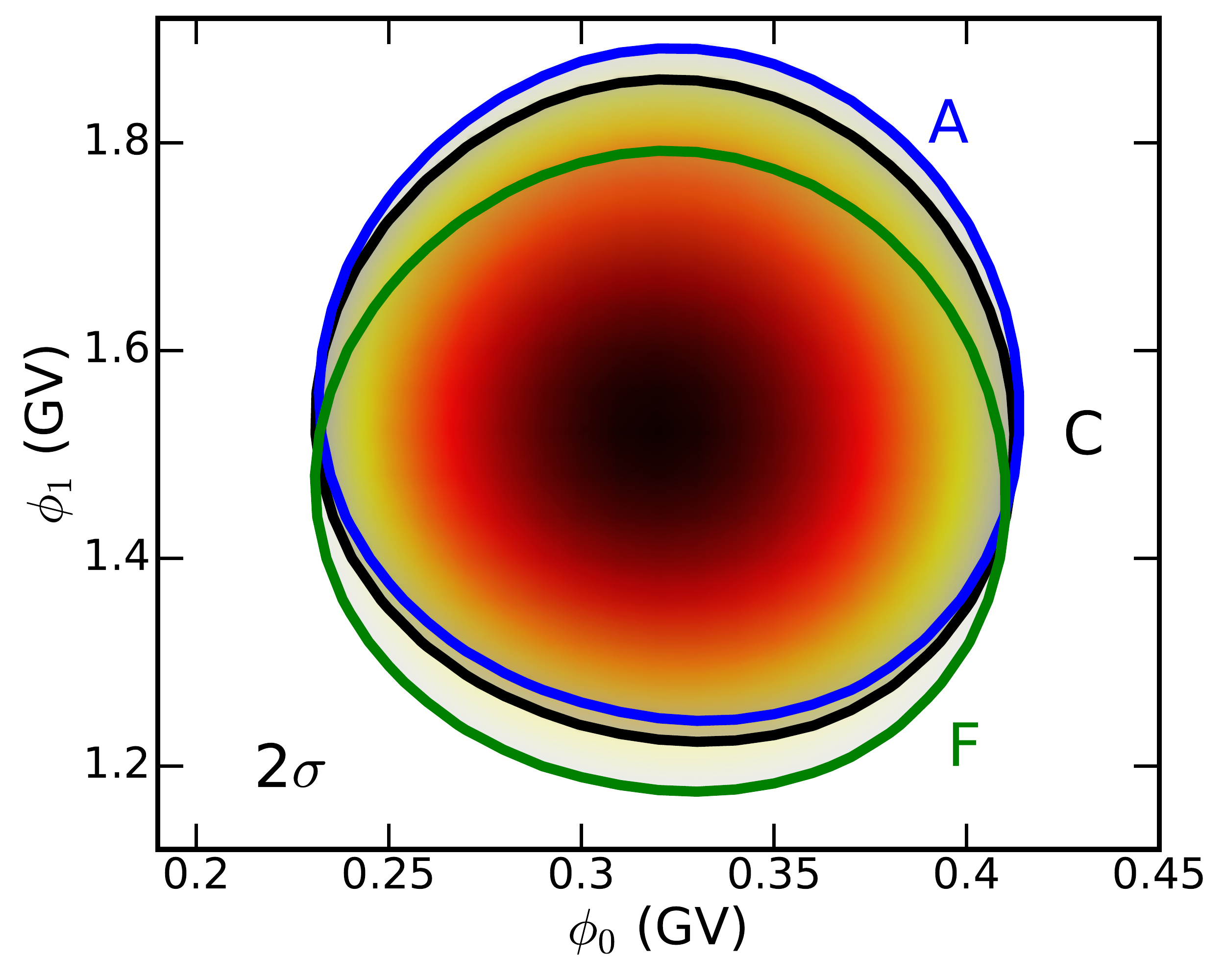}
\end{centering}
\caption{The 2$\sigma$ contours for the parameters $\phi_0$ and $\phi_1$, as found using models A, C, and F to describe the injection and transport of cosmic rays in the interstellar medium (see Table~\ref{tab:ISM_models}). Variations such as these have little impact on our results.}
\label{fig:PhiRangesISM}
\end{figure}

In Fig.~\ref{fig:PhiRanges}, we have also included the constraints on $\phi_0$ and $\phi_1$ that we have derived from the measurements of the antiproton-to-proton ratio, $\bar{p}/p$, by the combination of the BESS Polar II and \textit{AMS-02} experiments. In these fits, we have restricted our analysis to data above 0.4 GeV in kinetic energy, corresponding to approximately $R=1\,{\rm GV}$, below which our model is less reliable. We have also marginalized over the local gas density and the antiproton production cross section.  Although our constraints based on the antiproton-to-proton ratio are not very restrictive at this time, more sensitive measurements of the time-evolution of this ratio by \textit{AMS-02}, and the balloon-based GAPS experiment~\cite{Lowell:2018xff}, will increase the importance of this quantity. Furthermore, as future measurements reduce the uncertainties associated with the antiproton production cross section, the antiproton-to-proton ratio will become an increasingly powerful probe of solar modulation. In Appendix~\ref{sec:AppendixC}, we show the measured spectrum of the antiproton-to-proton ratio and compare this to the predictions of our model for solar modulation.

In Fig.~\ref{fig:PhiRangesISM}, we show constraints on $\phi_0$ and $\phi_1$, using three different models for cosmic-ray injection and transport in the ISM (models A, C, and F, from Table~\ref{tab:ISM_models}). This figure clearly demonstrates that such variations have little impact on our results. For the sake of clarity, we have not shown contours for models B, D, or E. We have tested these models, however, and found them to yield similar results, and favor values of $\phi_0$ and $\phi_1$ which fall within the ranges shown.

\begin{figure}
\begin{centering}
\hspace{-0.22in}
\includegraphics[width=3.65in,angle=0]{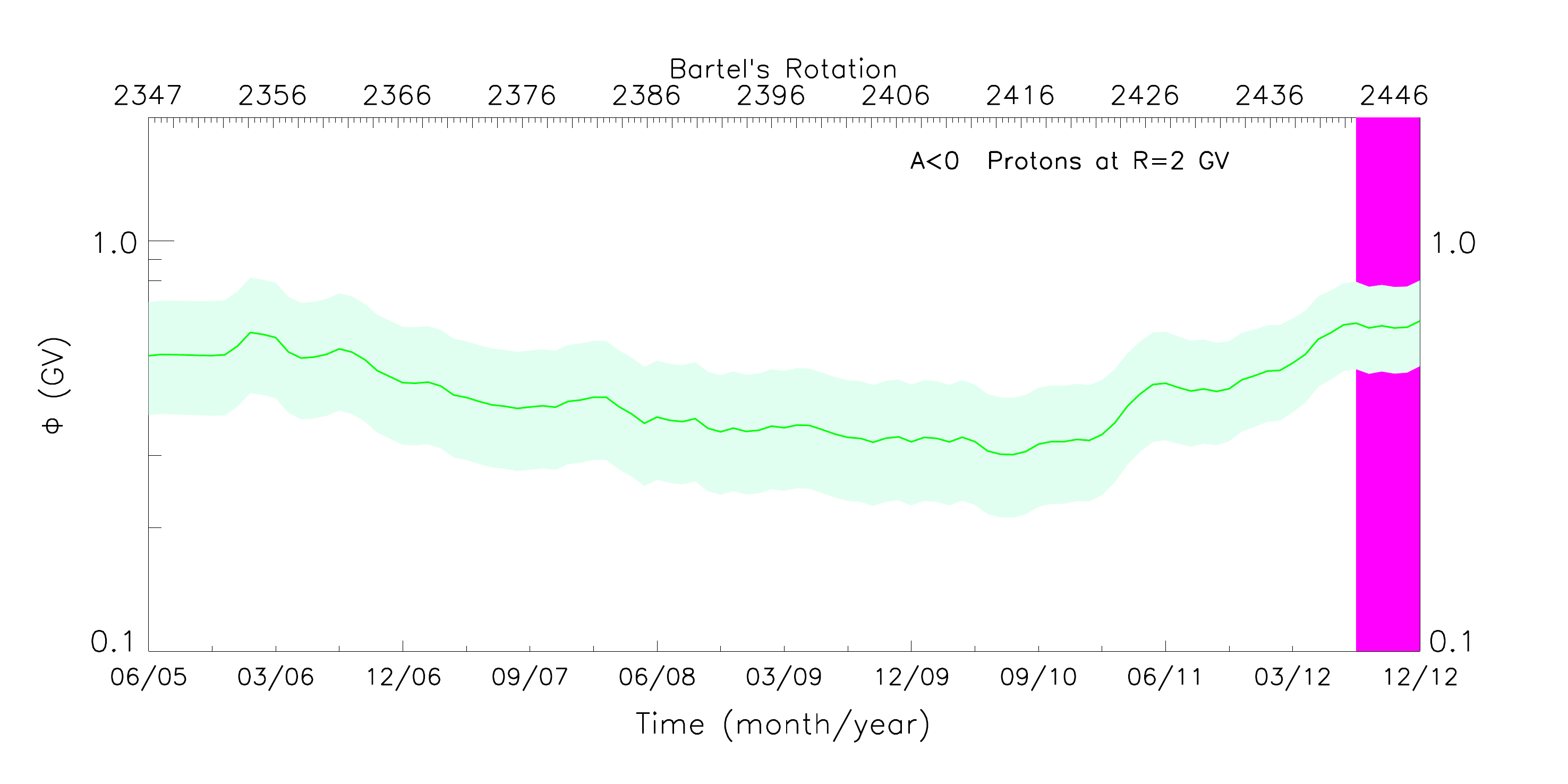}
\end{centering}
\vskip -0.25in
\caption{The time evolution of the solar modulation potential, $\Phi$, as evaluated for cosmic rays of charge $q=+1$ and at an ISM rigidity of $R=2\,{\rm GV}$. The green line shows the best-fit model, while the surrounding green band denotes the 3$\sigma$ uncertainty range. The vertical band at the far right denotes the period of the polarity flip of the heliospheric magnetic field, during which our model should not be applied.}
\label{fig:PhiTimeEvolution}
\end{figure}

Lastly, in Fig.~\ref{fig:PhiTimeEvolution}, we show the time evolution of the solar 
modulation potential, $\Phi$, during the time period considered in this study (June 2005-December 2012), evaluated for $q=+1$ and at a rigidity of $R=2 \, {\rm GV}$. Over this period of time, the modulation potential varies by approximately a factor of two.

\section{Conclusions}
\label{sec:conclusions}

In this study, we have used measurements of the cosmic-ray hydrogen spectrum from \textit{AMS-02}, and the antiproton-to-proton ratio from BESS Polar II, to model and constrain the effects of solar modulation on the cosmic-ray spectrum. By utilizing data from a period in which the heliospheric magnetic field was in a state of negative polarity, $A<0$, we have been able to observe and much more strongly constrain the charge-dependent effects of solar modulation.  

Building upon our previous work~\cite{Cholis:2015gna}, we have used the expression given in Eq.~\ref{eq:ModPot} to model the time-, charge- and rigidity-dependent effects of solar modulation. This model utilizes measurements of the heliospheric magnetic field and the tilt angle of the heliospheric current sheet to predict the impact of heliospheric forces on the cosmic-ray spectrum. Compared to a model with a constant solar modulation potential, our model improves the fit to the \textit{AMS-02} hydrogen spectrum at a level corresponding to $\Delta \chi^2 = 1160$, constituting an overwhelming statistical preference. We also find clear evidence that solar modulation impacts positively charged particles differently from those with a negative charge, supporting the conclusion that cosmic rays of opposite charge travel from the outer heliosphere to the Inner Solar System along different trajectories. More specifically, our results suggest that during periods of negative polarity, positively charged cosmic rays propagate from the heliopause into the Inner Solar System along the heliospheric current sheet, requiring about a year and leading to significant rigidity-dependent effects. In contrast, negatively charged particles are instead able to travel from the polar directions of the Solar System, reaching Earth more quickly, and experiencing less energy loss.

We note that our model allows one to accurately predict the effects of solar modulation for both negatively and positively charged cosmic rays as a function of time. This is a significant improvement over the force-field models that are typically employed in studies of cosmic-ray propagation. Moreover, the analytic nature of our results enables their usage in large Monte Carlo scans of the cosmic-ray diffusion parameter space. We thus strongly recommend the employment of these models as the standard treatment of solar modulation in cosmic-ray studies. To this end, we have produced a publicly available tool (\href{https://bitbucket.org/StockholmAstroparticle/solar-modulation/}{https://bitbucket.org/StockholmAstroparticle/solar-modulation/}), which allows one to straightforwardly and efficiently calculate the solar modulation parameters. This tool will be updated as future data and studies become available.

Our ability to accurately predict the effects of solar modulation is important for a variety of scientific goals, including efforts to isolate and understand the impact of convection and diffusive reacceleration on cosmic rays in the ISM. Additionally, the observed excess of $\sim$\,${\rm 10-20}$ GV cosmic-ray antiprotons has received attention as a possible signal of annihilating dark matter, and the systematic uncertainties related to the effects of solar modulation are one of the most important factors limiting our ability to confidently interpret this data~\cite{Cuoco:2016eej,Cui:2016ppb,Cholis:2019ejx,Boudaud:2019efq,Cuoco:2019kuu,Hooper:2019xss,Heisig:2020nse}.

\begin{table*}[t]
    \begin{tabular}{ccccccc}
         \hline
            $B_{\rm tot}(t)$ & $B_{\rm tot}(t)$ & $\alpha(t)$ & $\alpha(t)$ & $\phi_{0}$ (GV) & $\phi_{1}$ (GV) & $\Delta \chi^{2}$ \\
            time aver.  ($\#$BR) \, & time delay ($\#$BR) \, & time aver.  ($\#$BR) \, & time delay ($\#$BR) \, & 3$\sigma$ range \, & 3$\sigma$ range \, & from best fit  \\
            \hline \hline
            4  & 16 & 20 & 0 & 0.21-0.435 & 1.15-1.95 & -- \\
            4  & 17 & 21 & 0 & 0.13-0.32 & 1.15-1.85 & 46 \\
            4  & 17 & 4 & 17 & 0.06-0.245 & 1.15-1.70 & 59 \\
            4  & 16 & 4 & 16 & 0.12-0.35 & 0.95-1.60 & 61 \\
            7  & 0 & 7 & 0 & 0.59-0.88 & $\leq$0.70 & 108 \\
            7 & 0 & 4 & 0 & 0.605-0.88 & $\leq$0.95 & 108 \\
            6 & 0 & 3 & 0 & 0.60-0.90 & $\leq$0.75 & 117  \\
            4  & 18 & 4 & 18 & $\leq$0.155 & 1.40-2.15 & 125 \\
            4  & 15 & 4 & 15 & 0.135-0.36 & 0.85-1.45 & 127 \\
            23  & 0* & 13 & 0* & 0.585-0.88 & $\leq$0.30 & 151 \\
            4  & 18 & 22 & 0 & 0.07-0.23 & 1.20-1.90 & 163 \\
            9 & 0 & 6 & 0 & 0.57-0.84 & $\leq$0.45 & 203 \\
            10  & 0 & 10 & 0 & 0.535-0.78 & $\leq$0.25 & 237 \\
            13  & 0 & 13 & 0 & 0.515-0.75 & $\leq$0.30 & 253 \\
            13  & 0 & 7 & 0 & 0.515-0.75 & $\leq$0.30 & 253 \\
             7 & 1 & 7 & 0 & 0.505-0.76 & $\leq$0.75 & 257 \\
            18  & 0* & 9 & 0* & 0.58-0.89 & $\leq$0.45 & 348 \\
            3  & 0 & 3 & 0 & 0.315-0.54 & $\leq$1.00 & 573 \\
            1  & 0 & 1 & 0 & 0.17-0.30 & $\leq$0.25 & 785 \\                        
            \hline \hline 
        \end{tabular}
    \caption{A variety of schemes for time averaging the values of $B_{\rm tot}$ and $\alpha$, and the corresponding constraints on 
    $\phi_{0}$ and $\phi_{1}$ in each case. In the last column, we show the $\Delta \chi^{2}$ from the best fit
    choice. For the two models labeled with an asterisk, the averages were weighted as to favor more recent measurements of $B_{\rm tot}$ and $\alpha$, as described in the text.}
    \label{tab:Time_delay_fits}
\end{table*}                  
                  
\acknowledgements{IC acknowledges support from the Michigan Space Grant Consortium, under NASA Grant No.~NNX15AJ20H.
DH is supported by the Fermi Research Alliance, LLC, under Contract No.~DE-AC02-07CH11359 with the U.S. Department of Energy, Office of High Energy Physics. TL is partially supported by the Swedish Research Council under contract 2019-05135 and the Swedish National Space Agency under contract 117/19.}

\bibliography{SolMod_cycle23}

\begin{thebibliography}{30}
\expandafter\ifx\csname natexlab\endcsname\relax\def\natexlab#1{#1}\fi
\expandafter\ifx\csname bibnamefont\endcsname\relax
  \def\bibnamefont#1{#1}\fi
\expandafter\ifx\csname bibfnamefont\endcsname\relax
  \def\bibfnamefont#1{#1}\fi
\expandafter\ifx\csname citenamefont\endcsname\relax
  \def\citenamefont#1{#1}\fi
\expandafter\ifx\csname url\endcsname\relax
  \def\url#1{\texttt{#1}}\fi
\expandafter\ifx\csname urlprefix\endcsname\relax\def\urlprefix{URL }\fi
\providecommand{\bibinfo}[2]{#2}
\providecommand{\eprint}[2][]{\url{#2}}

\bibitem[{\citenamefont{Cholis et~al.}(2016)\citenamefont{Cholis, Hooper, and
  Linden}}]{Cholis:2015gna}
\bibinfo{author}{\bibfnamefont{I.}~\bibnamefont{Cholis}},
  \bibinfo{author}{\bibfnamefont{D.}~\bibnamefont{Hooper}}, \bibnamefont{and}
  \bibinfo{author}{\bibfnamefont{T.}~\bibnamefont{Linden}},
  \bibinfo{journal}{Phys. Rev.} \textbf{\bibinfo{volume}{D93}},
  \bibinfo{pages}{043016} (\bibinfo{year}{2016}), \eprint{1511.01507}.

\bibitem[{\citenamefont{Corti et~al.}(2020)\citenamefont{Corti, Bindi,
  Consolandi, Freeman, Kuhlman, Light, Palermo, and Wang}}]{Corti:2019jlt}
\bibinfo{author}{\bibfnamefont{C.}~\bibnamefont{Corti}},
  \bibinfo{author}{\bibfnamefont{V.}~\bibnamefont{Bindi}},
  \bibinfo{author}{\bibfnamefont{C.}~\bibnamefont{Consolandi}},
  \bibinfo{author}{\bibfnamefont{C.}~\bibnamefont{Freeman}},
  \bibinfo{author}{\bibfnamefont{A.}~\bibnamefont{Kuhlman}},
  \bibinfo{author}{\bibfnamefont{C.}~\bibnamefont{Light}},
  \bibinfo{author}{\bibfnamefont{M.}~\bibnamefont{Palermo}}, \bibnamefont{and}
  \bibinfo{author}{\bibfnamefont{S.}~\bibnamefont{Wang}},
  \bibinfo{journal}{PoS} \textbf{\bibinfo{volume}{ICRC2019}},
  \bibinfo{pages}{1070} (\bibinfo{year}{2020}), \eprint{1910.00027}.

\bibitem[{\citenamefont{Boschini et~al.}(2019)\citenamefont{Boschini,
  Della~Torre, Gervasi, La~Vacca, and Rancoita}}]{Boschini:2019ubh}
\bibinfo{author}{\bibfnamefont{M.~J.} \bibnamefont{Boschini}},
  \bibinfo{author}{\bibfnamefont{S.}~\bibnamefont{Della~Torre}},
  \bibinfo{author}{\bibfnamefont{M.}~\bibnamefont{Gervasi}},
  \bibinfo{author}{\bibfnamefont{G.}~\bibnamefont{La~Vacca}}, \bibnamefont{and}
  \bibinfo{author}{\bibfnamefont{P.~G.} \bibnamefont{Rancoita}},
  \bibinfo{journal}{Adv. Space Res.} \textbf{\bibinfo{volume}{64}},
  \bibinfo{pages}{2459} (\bibinfo{year}{2019}), \eprint{1903.07501}.

\bibitem[{\citenamefont{Wang et~al.}(2019)\citenamefont{Wang, Bi, Fang, Lin,
  and Yin}}]{Wang:2019xtu}
\bibinfo{author}{\bibfnamefont{B.-B.} \bibnamefont{Wang}},
  \bibinfo{author}{\bibfnamefont{X.-J.} \bibnamefont{Bi}},
  \bibinfo{author}{\bibfnamefont{K.}~\bibnamefont{Fang}},
  \bibinfo{author}{\bibfnamefont{S.-J.} \bibnamefont{Lin}}, \bibnamefont{and}
  \bibinfo{author}{\bibfnamefont{P.-F.} \bibnamefont{Yin}},
  \bibinfo{journal}{Phys. Rev. D} \textbf{\bibinfo{volume}{100}},
  \bibinfo{pages}{063006} (\bibinfo{year}{2019}), \eprint{1904.03747}.

\bibitem[{\citenamefont{Gieseler et~al.}(2017)\citenamefont{Gieseler, Heber,
  and Herbst}}]{Gieseler:2017xry}
\bibinfo{author}{\bibfnamefont{J.}~\bibnamefont{Gieseler}},
  \bibinfo{author}{\bibfnamefont{B.}~\bibnamefont{Heber}}, \bibnamefont{and}
  \bibinfo{author}{\bibfnamefont{K.}~\bibnamefont{Herbst}},
  \bibinfo{journal}{J. Geophys. Res. Space Phys.}
  \textbf{\bibinfo{volume}{122}}, \bibinfo{pages}{10,964}
  (\bibinfo{year}{2017}), \eprint{1710.10834}.

\bibitem[{\citenamefont{Corti et~al.}(2019)\citenamefont{Corti, Potgieter,
  Bindi, Consolandi, Light, Palermo, and Popkow}}]{Corti:2018ycg}
\bibinfo{author}{\bibfnamefont{C.}~\bibnamefont{Corti}},
  \bibinfo{author}{\bibfnamefont{M.~S.} \bibnamefont{Potgieter}},
  \bibinfo{author}{\bibfnamefont{V.}~\bibnamefont{Bindi}},
  \bibinfo{author}{\bibfnamefont{C.}~\bibnamefont{Consolandi}},
  \bibinfo{author}{\bibfnamefont{C.}~\bibnamefont{Light}},
  \bibinfo{author}{\bibfnamefont{M.}~\bibnamefont{Palermo}}, \bibnamefont{and}
  \bibinfo{author}{\bibfnamefont{A.}~\bibnamefont{Popkow}},
  \bibinfo{journal}{Astrophys. J.} \textbf{\bibinfo{volume}{871}},
  \bibinfo{pages}{253} (\bibinfo{year}{2019}), \eprint{1810.09640}.

\bibitem[{\citenamefont{Aslam et~al.}(2019)\citenamefont{Aslam, Bisschoff,
  Potgieter, Boezio, and Munini}}]{Aslam:2018kpi}
\bibinfo{author}{\bibfnamefont{O.}~\bibnamefont{Aslam}},
  \bibinfo{author}{\bibfnamefont{D.}~\bibnamefont{Bisschoff}},
  \bibinfo{author}{\bibfnamefont{M.}~\bibnamefont{Potgieter}},
  \bibinfo{author}{\bibfnamefont{M.}~\bibnamefont{Boezio}}, \bibnamefont{and}
  \bibinfo{author}{\bibfnamefont{R.}~\bibnamefont{Munini}},
  \bibinfo{journal}{Astrophys. J.} \textbf{\bibinfo{volume}{873}},
  \bibinfo{pages}{70} (\bibinfo{year}{2019}), \eprint{1811.10710}.

\bibitem[{\citenamefont{Bertucci et~al.}(2020)\citenamefont{Bertucci,
  Fiandrini, Khiali, and Tomassetti}}]{Bertucci:2019ypo}
\bibinfo{author}{\bibfnamefont{B.}~\bibnamefont{Bertucci}},
  \bibinfo{author}{\bibfnamefont{E.}~\bibnamefont{Fiandrini}},
  \bibinfo{author}{\bibfnamefont{B.}~\bibnamefont{Khiali}}, \bibnamefont{and}
  \bibinfo{author}{\bibfnamefont{N.}~\bibnamefont{Tomassetti}},
  \bibinfo{journal}{PoS} \textbf{\bibinfo{volume}{ICRC2019}},
  \bibinfo{pages}{1161} (\bibinfo{year}{2020}), \eprint{1908.01599}.

\bibitem[{\citenamefont{{Gleeson} and {Axford}}(1968)}]{1968ApJ...154.1011G}
\bibinfo{author}{\bibfnamefont{L.~J.} \bibnamefont{{Gleeson}}}
  \bibnamefont{and} \bibinfo{author}{\bibfnamefont{W.~I.}
  \bibnamefont{{Axford}}}, \bibinfo{journal}{\apj}
  \textbf{\bibinfo{volume}{154}}, \bibinfo{pages}{1011} (\bibinfo{year}{1968}).

\bibitem[{\citenamefont{{Strauss} et~al.}(2012)\citenamefont{{Strauss},
  {Potgieter}, {B{\"u}sching}, and {Kopp}}}]{2012Ap&SS.339..223S}
\bibinfo{author}{\bibfnamefont{R.~D.} \bibnamefont{{Strauss}}},
  \bibinfo{author}{\bibfnamefont{M.~S.} \bibnamefont{{Potgieter}}},
  \bibinfo{author}{\bibfnamefont{I.}~\bibnamefont{{B{\"u}sching}}},
  \bibnamefont{and} \bibinfo{author}{\bibfnamefont{A.}~\bibnamefont{{Kopp}}},
  \bibinfo{journal}{Astro.~and Space Sci.} \textbf{\bibinfo{volume}{339}},
  \bibinfo{pages}{223} (\bibinfo{year}{2012}).

\bibitem[{\citenamefont{Potgieter}(2013)}]{Potgieter:2013pdj}
\bibinfo{author}{\bibfnamefont{M.}~\bibnamefont{Potgieter}},
  \bibinfo{journal}{Living Rev. Solar Phys.} \textbf{\bibinfo{volume}{10}},
  \bibinfo{pages}{3} (\bibinfo{year}{2013}), \eprint{1306.4421}.

\bibitem[{\citenamefont{Consolandi}(2016)}]{Consolandi:2015sop}
\bibinfo{author}{\bibfnamefont{C.}~\bibnamefont{Consolandi}},
  \bibinfo{journal}{PoS} \textbf{\bibinfo{volume}{ICRC2015}},
  \bibinfo{pages}{117} (\bibinfo{year}{2016}).

\bibitem[{\citenamefont{Abe et~al.}(2017)}]{Abe:2017yrg}
\bibinfo{author}{\bibfnamefont{K.}~\bibnamefont{Abe}} \bibnamefont{et~al.},
  \bibinfo{journal}{Adv. Space Res.} \textbf{\bibinfo{volume}{60}},
  \bibinfo{pages}{806} (\bibinfo{year}{2017}).

\bibitem[{\citenamefont{http://www.srl.caltech.edu/ACE/ASC/}()}]{ACESite}
\bibinfo{author}{\bibnamefont{http://www.srl.caltech.edu/ACE/ASC/}}.

\bibitem[{\citenamefont{http://wso.stanford.edu/Tilts.html}()}]{WSOSite}
\bibinfo{author}{\bibnamefont{http://wso.stanford.edu/Tilts.html}}.

\bibitem[{\citenamefont{http://galprop.stanford.edu/.}()}]{GALPROPSite}
\bibinfo{author}{\bibnamefont{http://galprop.stanford.edu/.}}

\bibitem[{\citenamefont{Strong}(2015)}]{Strong:2015zva}
\bibinfo{author}{\bibfnamefont{A.~W.} \bibnamefont{Strong}}
  (\bibinfo{year}{2015}), \eprint{1507.05020}.

\bibitem[{\citenamefont{version of GALPROP availabe~at:
  http://sourceforge.net/projects/galprop}()}]{NEWGALPROP}
\bibinfo{author}{\bibfnamefont{N.}~\bibnamefont{version of GALPROP availabe~at:
  http://sourceforge.net/projects/galprop}}.

\bibitem[{\citenamefont{Strong and Moskalenko}(1998)}]{galprop}
\bibinfo{author}{\bibfnamefont{A.~W.} \bibnamefont{Strong}} \bibnamefont{and}
  \bibinfo{author}{\bibfnamefont{I.~V.} \bibnamefont{Moskalenko}},
  \bibinfo{journal}{Astrophys. J.} \textbf{\bibinfo{volume}{509}},
  \bibinfo{pages}{212} (\bibinfo{year}{1998}), \eprint{astro-ph/9807150}.

\bibitem[{\citenamefont{Cholis et~al.}(2017)\citenamefont{Cholis, Hooper, and
  Linden}}]{Cholis:2017qlb}
\bibinfo{author}{\bibfnamefont{I.}~\bibnamefont{Cholis}},
  \bibinfo{author}{\bibfnamefont{D.}~\bibnamefont{Hooper}}, \bibnamefont{and}
  \bibinfo{author}{\bibfnamefont{T.}~\bibnamefont{Linden}},
  \bibinfo{journal}{Phys. Rev.} \textbf{\bibinfo{volume}{D95}},
  \bibinfo{pages}{123007} (\bibinfo{year}{2017}), \eprint{1701.04406}.

\bibitem[{\citenamefont{Cholis et~al.}(2019)\citenamefont{Cholis, Linden, and
  Hooper}}]{Cholis:2019ejx}
\bibinfo{author}{\bibfnamefont{I.}~\bibnamefont{Cholis}},
  \bibinfo{author}{\bibfnamefont{T.}~\bibnamefont{Linden}}, \bibnamefont{and}
  \bibinfo{author}{\bibfnamefont{D.}~\bibnamefont{Hooper}},
  \bibinfo{journal}{Phys. Rev.} \textbf{\bibinfo{volume}{D99}},
  \bibinfo{pages}{103026} (\bibinfo{year}{2019}), \eprint{1903.02549}.

\bibitem[{\citenamefont{Lowell et~al.}(2019)\citenamefont{Lowell, Aramaki, and
  Bird}}]{Lowell:2018xff}
\bibinfo{author}{\bibfnamefont{A.}~\bibnamefont{Lowell}},
  \bibinfo{author}{\bibfnamefont{T.}~\bibnamefont{Aramaki}}, \bibnamefont{and}
  \bibinfo{author}{\bibfnamefont{R.}~\bibnamefont{Bird}}
  (\bibinfo{collaboration}{GAPS}), \bibinfo{journal}{PoS}
  \textbf{\bibinfo{volume}{ICHEP2018}}, \bibinfo{pages}{543}
  (\bibinfo{year}{2019}), \eprint{1812.04800}.

\bibitem[{\citenamefont{Cuoco et~al.}(2017)\citenamefont{Cuoco, Krämer, and
  Korsmeier}}]{Cuoco:2016eej}
\bibinfo{author}{\bibfnamefont{A.}~\bibnamefont{Cuoco}},
  \bibinfo{author}{\bibfnamefont{M.}~\bibnamefont{Krämer}}, \bibnamefont{and}
  \bibinfo{author}{\bibfnamefont{M.}~\bibnamefont{Korsmeier}},
  \bibinfo{journal}{Phys. Rev. Lett.} \textbf{\bibinfo{volume}{118}},
  \bibinfo{pages}{191102} (\bibinfo{year}{2017}), \eprint{1610.03071}.

\bibitem[{\citenamefont{Cui et~al.}(2017)\citenamefont{Cui, Yuan, Tsai, and
  Fan}}]{Cui:2016ppb}
\bibinfo{author}{\bibfnamefont{M.-Y.} \bibnamefont{Cui}},
  \bibinfo{author}{\bibfnamefont{Q.}~\bibnamefont{Yuan}},
  \bibinfo{author}{\bibfnamefont{Y.-L.~S.} \bibnamefont{Tsai}},
  \bibnamefont{and} \bibinfo{author}{\bibfnamefont{Y.-Z.} \bibnamefont{Fan}},
  \bibinfo{journal}{Phys. Rev. Lett.} \textbf{\bibinfo{volume}{118}},
  \bibinfo{pages}{191101} (\bibinfo{year}{2017}), \eprint{1610.03840}.

\bibitem[{\citenamefont{Boudaud et~al.}(2020)\citenamefont{Boudaud, Génolini,
  Derome, Lavalle, Maurin, Salati, and Serpico}}]{Boudaud:2019efq}
\bibinfo{author}{\bibfnamefont{M.}~\bibnamefont{Boudaud}},
  \bibinfo{author}{\bibfnamefont{Y.}~\bibnamefont{Génolini}},
  \bibinfo{author}{\bibfnamefont{L.}~\bibnamefont{Derome}},
  \bibinfo{author}{\bibfnamefont{J.}~\bibnamefont{Lavalle}},
  \bibinfo{author}{\bibfnamefont{D.}~\bibnamefont{Maurin}},
  \bibinfo{author}{\bibfnamefont{P.}~\bibnamefont{Salati}}, \bibnamefont{and}
  \bibinfo{author}{\bibfnamefont{P.~D.} \bibnamefont{Serpico}},
  \bibinfo{journal}{Phys. Rev. Res.} \textbf{\bibinfo{volume}{2}},
  \bibinfo{pages}{023022} (\bibinfo{year}{2020}), \eprint{1906.07119}.

\bibitem[{\citenamefont{Cuoco et~al.}(2019)\citenamefont{Cuoco, Heisig, Klamt,
  Korsmeier, and Krämer}}]{Cuoco:2019kuu}
\bibinfo{author}{\bibfnamefont{A.}~\bibnamefont{Cuoco}},
  \bibinfo{author}{\bibfnamefont{J.}~\bibnamefont{Heisig}},
  \bibinfo{author}{\bibfnamefont{L.}~\bibnamefont{Klamt}},
  \bibinfo{author}{\bibfnamefont{M.}~\bibnamefont{Korsmeier}},
  \bibnamefont{and} \bibinfo{author}{\bibfnamefont{M.}~\bibnamefont{Krämer}},
  \bibinfo{journal}{Phys. Rev. D} \textbf{\bibinfo{volume}{99}},
  \bibinfo{pages}{103014} (\bibinfo{year}{2019}), \eprint{1903.01472}.

\bibitem[{\citenamefont{Hooper et~al.}(2019)\citenamefont{Hooper, Leane, Tsai,
  Wegsman, and Witte}}]{Hooper:2019xss}
\bibinfo{author}{\bibfnamefont{D.}~\bibnamefont{Hooper}},
  \bibinfo{author}{\bibfnamefont{R.~K.} \bibnamefont{Leane}},
  \bibinfo{author}{\bibfnamefont{Y.-D.} \bibnamefont{Tsai}},
  \bibinfo{author}{\bibfnamefont{S.}~\bibnamefont{Wegsman}}, \bibnamefont{and}
  \bibinfo{author}{\bibfnamefont{S.~J.} \bibnamefont{Witte}}
  (\bibinfo{year}{2019}), \eprint{1912.08821}.

\bibitem[{\citenamefont{Heisig et~al.}(2020)\citenamefont{Heisig, Korsmeier,
  and Winkler}}]{Heisig:2020nse}
\bibinfo{author}{\bibfnamefont{J.}~\bibnamefont{Heisig}},
  \bibinfo{author}{\bibfnamefont{M.}~\bibnamefont{Korsmeier}},
  \bibnamefont{and} \bibinfo{author}{\bibfnamefont{M.~W.}
  \bibnamefont{Winkler}} (\bibinfo{year}{2020}), \eprint{2005.04237}.

\bibitem[{\citenamefont{Aguilar et~al.}(2016)}]{Aguilar:2016kjl}
\bibinfo{author}{\bibfnamefont{M.}~\bibnamefont{Aguilar}} \bibnamefont{et~al.}
  (\bibinfo{collaboration}{AMS}), \bibinfo{journal}{Phys. Rev. Lett.}
  \textbf{\bibinfo{volume}{117}}, \bibinfo{pages}{091103}
  (\bibinfo{year}{2016}).

\bibitem[{\citenamefont{di~Mauro et~al.}(2014)\citenamefont{di~Mauro, Donato,
  Goudelis, and Serpico}}]{diMauro:2014zea}
\bibinfo{author}{\bibfnamefont{M.}~\bibnamefont{di~Mauro}},
  \bibinfo{author}{\bibfnamefont{F.}~\bibnamefont{Donato}},
  \bibinfo{author}{\bibfnamefont{A.}~\bibnamefont{Goudelis}}, \bibnamefont{and}
  \bibinfo{author}{\bibfnamefont{P.~D.} \bibnamefont{Serpico}},
  \bibinfo{journal}{Phys. Rev.} \textbf{\bibinfo{volume}{D90}},
  \bibinfo{pages}{085017} (\bibinfo{year}{2014}), \eprint{1408.0288}.

\end{thebibliography}
\bibliographystyle{apsrev}

\newpage

\begin{appendix}

\section{Time Averaging the Heliospheric Magnetic Field and Current Sheet Tilt Angle}
\label{sec:AppendixB}

In Table~\ref{tab:Time_delay_fits}, we show our constraints on $\phi_0$ and $\phi_1$, for a variety of schemes for time-averaging the values of $B_{\rm tot}$ and $\alpha$. For each case, we also show the total change to the overall quality of the fit to the {\textit AMS-02} hydrogen data. We have considered cases with time delays as large as 18 Bartles' Rotations, and averaging windows as large as 23 Bartles' Rotations.

\begin{figure}
\begin{centering}
\includegraphics[width=3.25in,angle=0]{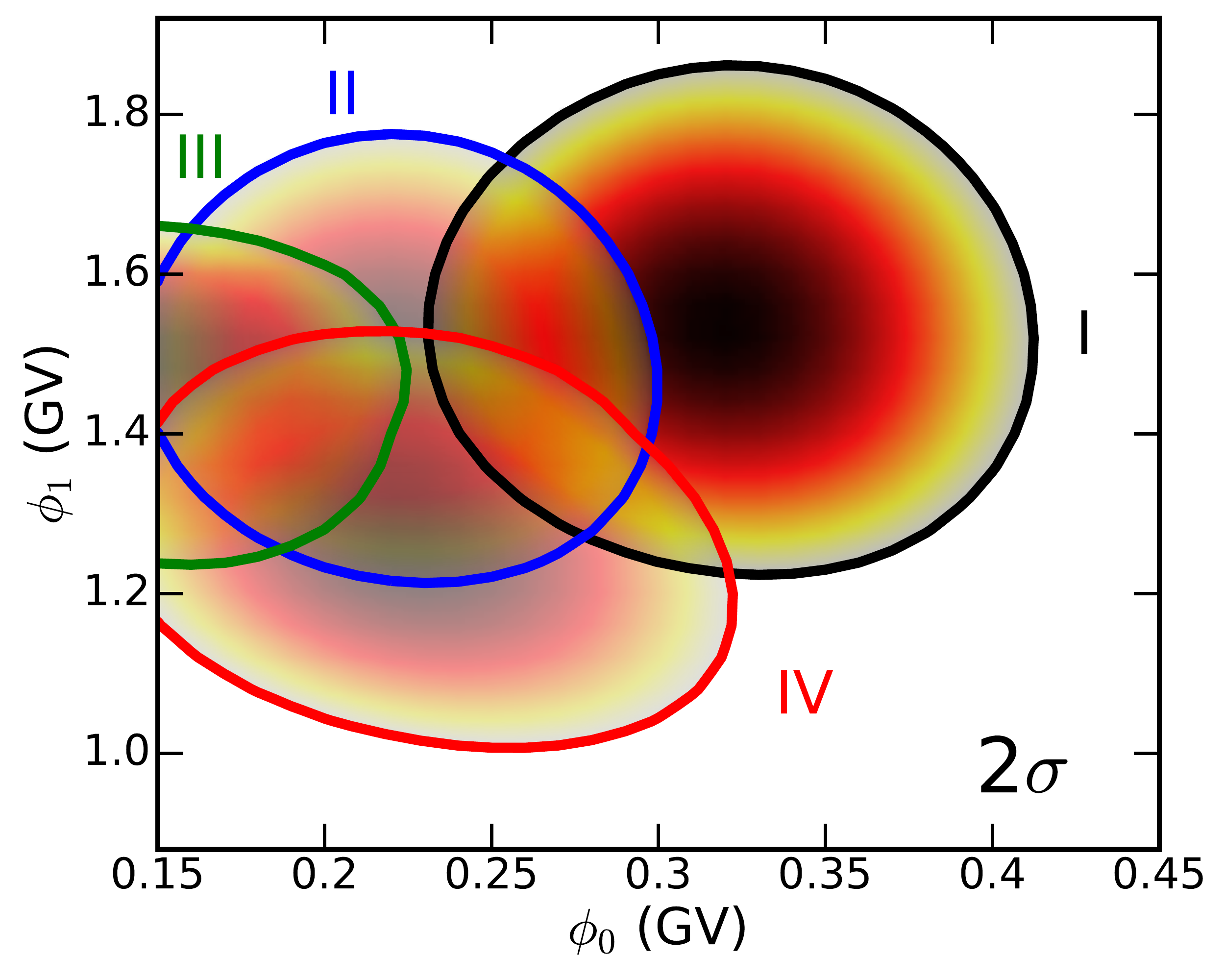}
\end{centering}
\caption{The 2$\sigma$ confidence contours on $\phi_0$ and $\phi_1$ for the four best-fit schemes for the time averaging of the values of $B_{\rm tot}$ and $\alpha$. The labels I, II, III and IV correspond to the 1st, 2nd, 3rd and 4th rows of Table~\ref{tab:Time_delay_fits}, respectively.}
\label{fig:PhiRangesTime}
\end{figure}

As previously stated in the main text, the best fit is found when $B_{\rm tot}$ is averaged over four Bartels' Rotations, 16 to 19 rotations prior to the time of the cosmic-ray measurement. Similarly, the best fit is found when we average $\alpha$ over the past 20 Bartles' Rotations. Other choices pertaining to the averaging of these quantities yield fits that are often significantly worse, as shown in Table~\ref{tab:Time_delay_fits}). In Fig.~\ref{fig:PhiRangesTime}, we show the favored ranges of $\phi_0$ and $\phi_1$ for the cases corresponding to the top four rows in Table~\ref{tab:Time_delay_fits}. Note that in all four of these cases, the data strongly prefer a non-zero value of $\phi_1$.

In most of the scenarios we have considered, we have applied equal weights to each Bartels' Rotation included in the average. As a well-motivated alternative, we have also considered two scenarios with no time delay, but with greater weighting applied to the most recent measurements of $B_{\rm tot}$ and $\alpha$. In particular, in 10th line of Table~\ref{tab:Time_delay_fits}, we have adopted weights of $1/(i+15)$ for $B_{\rm tot}$ and $1/(i+8.26)$ for $\alpha$, where $i$ is number of Bartles' Rotations in the past that a given measurement was taken. A similar weighting scheme was applied to the model shown in the 17th line of the Table. Neither of these cases were found to be favored by the data.

\bigskip

\section{The Antiproton-to-Proton Ratio}
\label{sec:AppendixC}

\begin{figure*}
\begin{centering}
\includegraphics[width=3.68in,angle=0]{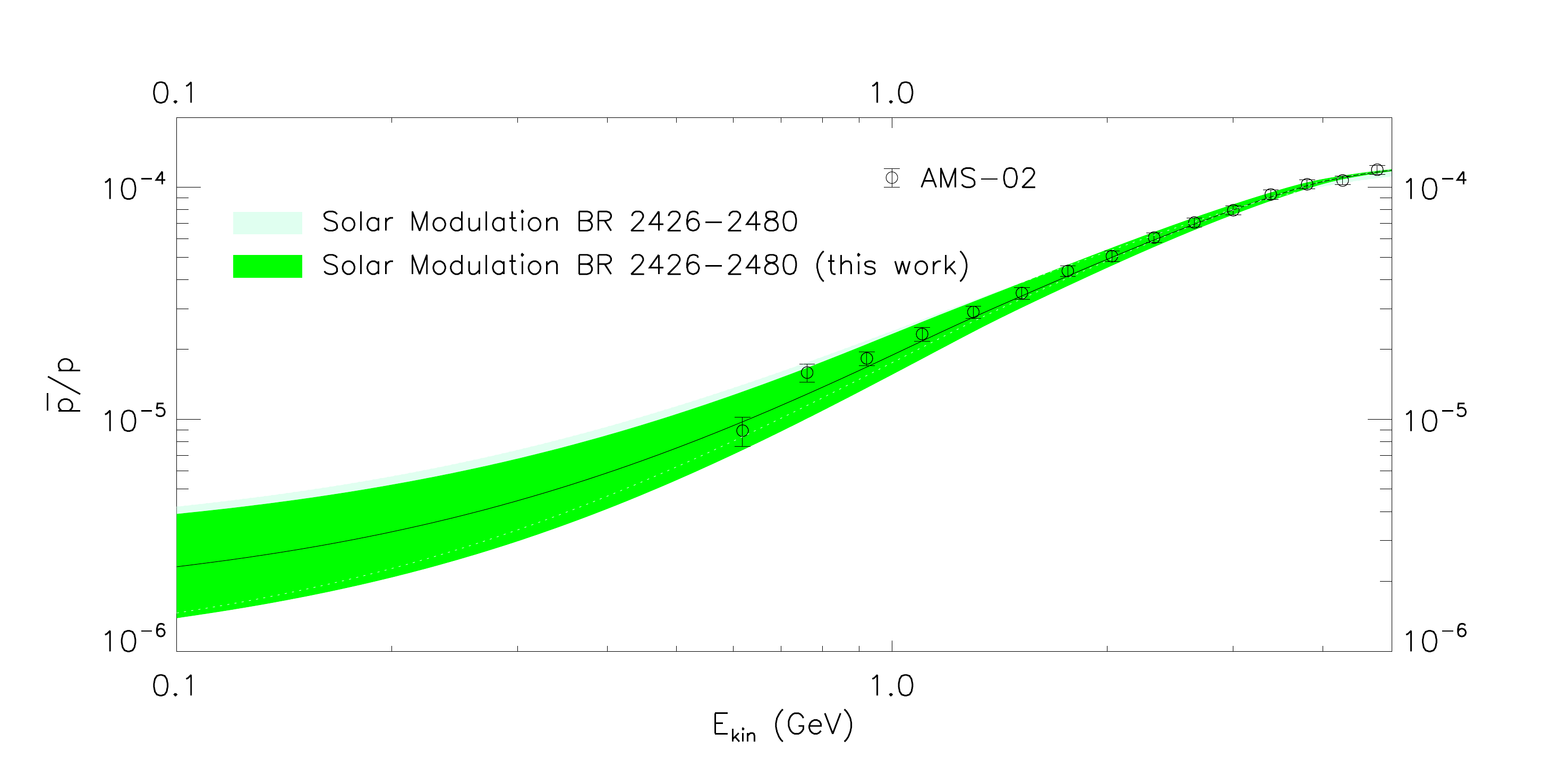}
\hspace{-1.0cm}
\includegraphics[width=3.68in,angle=0]{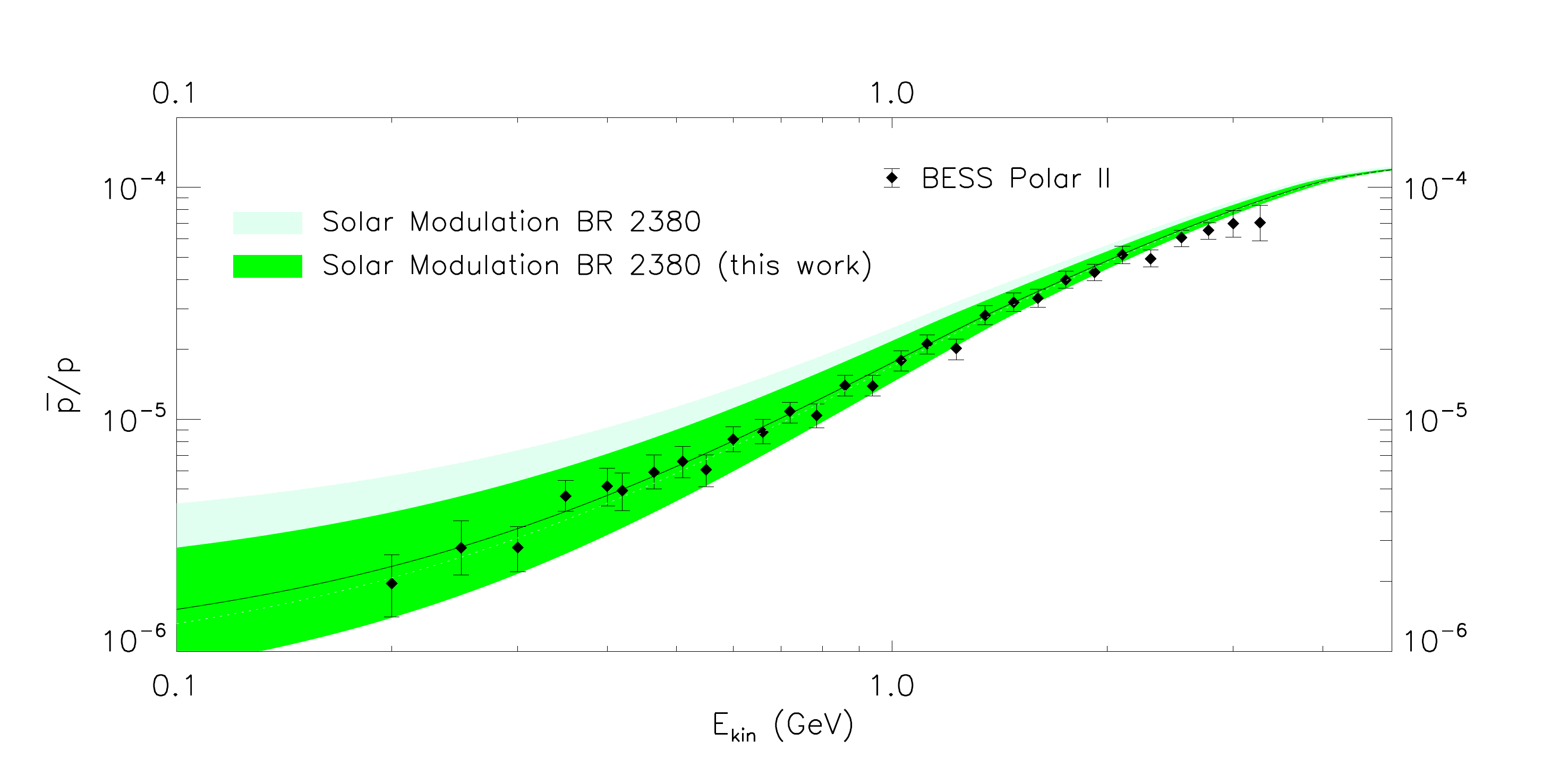}  \\
\includegraphics[width=3.68in,angle=0]{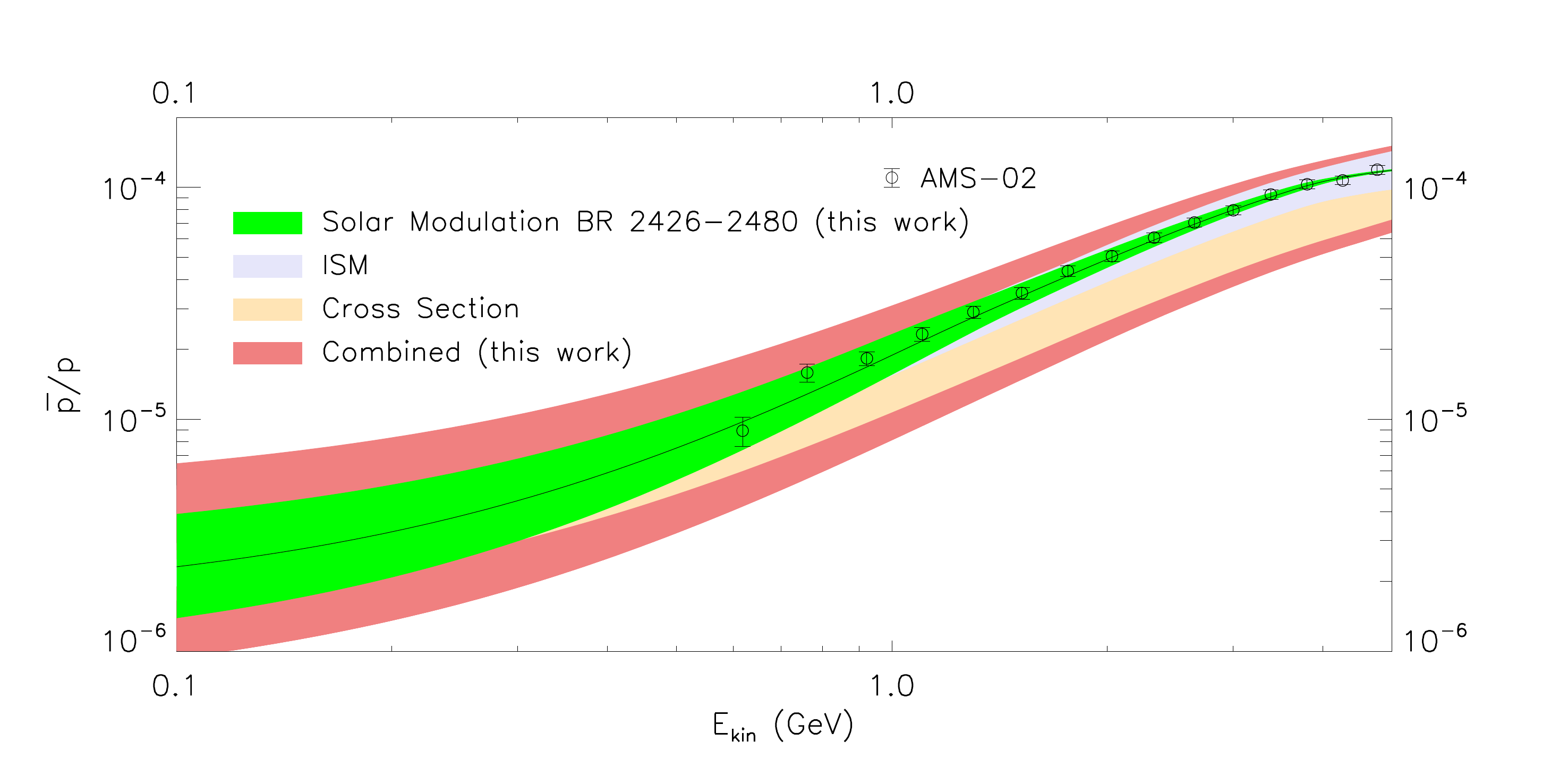} 
\hspace{-1.0cm}
\includegraphics[width=3.68in,angle=0]{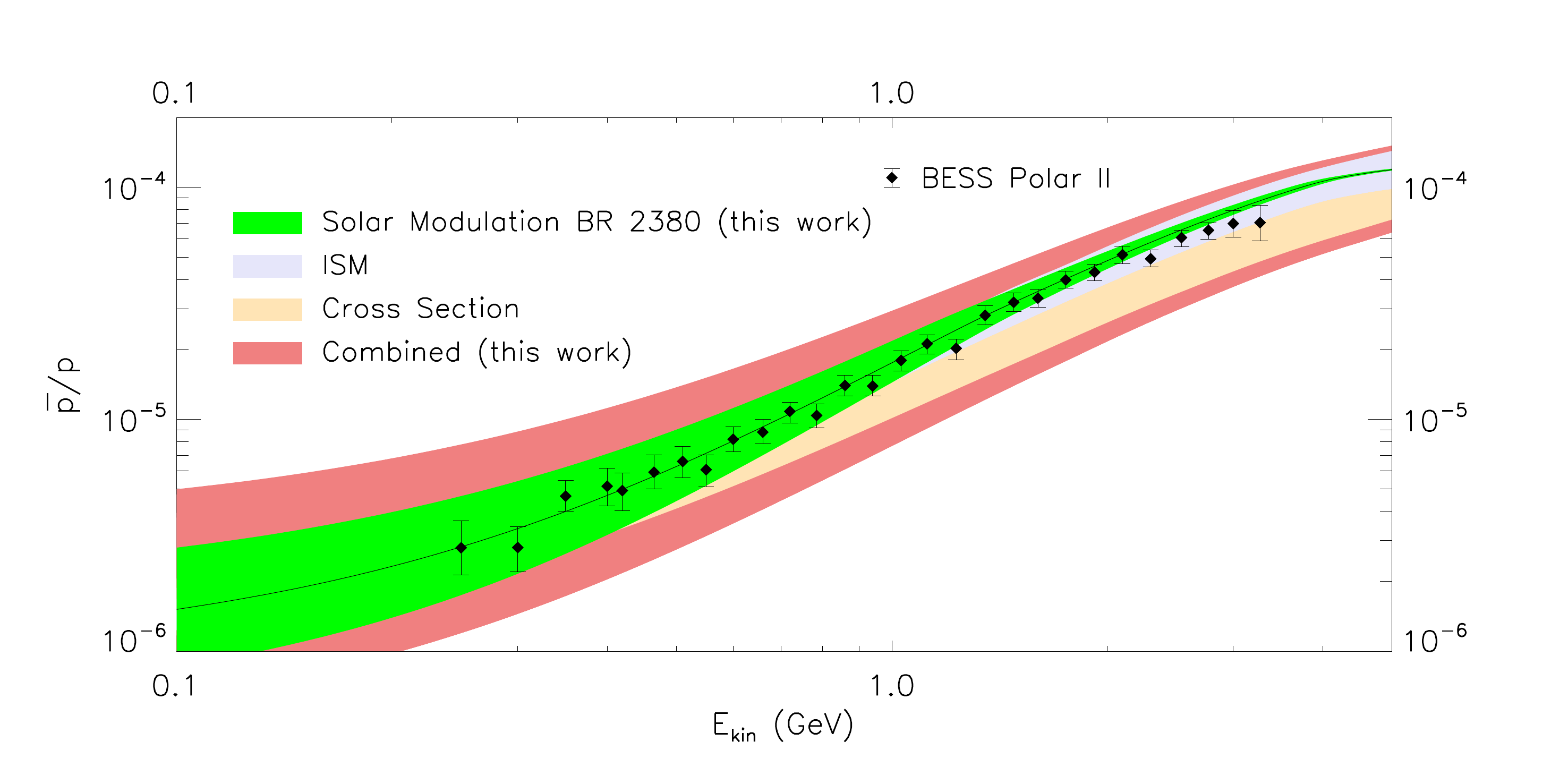} 
\end{centering}
\caption{The cosmic-ray antiproton-to-proton ratio as measured by \textit{AMS-02}\cite{Aguilar:2016kjl} (left frames) and by BESS Polar II (right frames). These measurements are compared to the predictions of our model (solid curves), including the uncertainties associated with solar modulation. In each of the top frames, we show the uncertainties related to solar modulation using the constraints on $\phi_0$ and $\phi_1$ from our previous work~\cite{Cholis:2015gna} (larger bands), and as constrained in this study (smaller bands). In the lower frames, we include the current uncertainties associated with solar modulation, as well as those associated with the injection and propagation in the ISM, and the antiproton production cross section~\cite{Cholis:2017qlb,diMauro:2014zea}. For the BESS Polar II data, we show the spectrum as predicted during Bartels' Rotation 2380 (December 2007), while for \textit{AMS-02} we show the spectrum of this ratio averaged over the period of the dataset shown.}
\label{fig:PbarToP}
\end{figure*}

In Fig.~\ref{fig:PbarToP}, we show the cosmic-ray antiproton-to-proton ratio as measured by \textit{AMS-02}~\cite{Aguilar:2016kjl} (left frames) and by BESS Polar II~\cite{Abe:2017yrg} (right frames), and compare these results to the prediction of our model, including the uncertainties associated with solar modulation. In each of the top frames, we show the uncertainties related to solar modulation using the constraints on $\phi_0$ and $\phi_1$ from our previous work~\cite{Cholis:2015gna}, and as constrained in this study. In the lower frames, we include the current uncertainties associated with solar modulation, as well as those associated with the injection and propagation in the ISM (including variations in the local gas density), and the antiproton production cross section~\cite{Cholis:2017qlb,diMauro:2014zea}. For the BESS Polar II data, we show the spectrum as predicted during Bartels' Rotation 2380 (December 2007), while for \textit{AMS-02} we show the spectrum of this ratio averaged over the period of the dataset shown.

\end{appendix}  
                  
\end{document}